\documentclass[aps,pra,reprint,superscriptaddress]{revtex4-1}
\usepackage{CJK}
\usepackage{xcolor}
\usepackage{amsmath}
\usepackage{amssymb}
\usepackage{hyperref}
\usepackage{cleveref}
\usepackage[arrowdel]{physics}
\usepackage{dsfont}

\usepackage{MnSymbol}

\usepackage{soul}
\usepackage{cancel}

\usepackage{graphicx}
\usepackage{subfig}
\graphicspath{{figures/}}

\newcounter{subeqn}
\renewcommand{\thesubeqn}{\theequation\alph{subeqn}}%
\newcommand{\subeqn}{%
  \refstepcounter{subeqn}
  \tag{\thesubeqn}
}

\newcommand{\beginsubeqn}{\setcounter{subeqn}{0}\refstepcounter{equation}\subeqn}

\newcommand{\subfigimg}[4][,]{%
	\setbox1=\hbox{\includegraphics[#1]{#2}}
	\leavevmode\rlap{\usebox1}
	\rlap{\hspace*{#3}\raisebox{\dimexpr\ht1-#4\baselineskip}{\fcolorbox{black}{white}{(\alph{subfigure})}}}
	\phantom{\usebox1}
}

\DeclareDocumentCommand\bbra{ s m t\kket s g }
{ 
	\IfBooleanTF{#3}
	{ 
		\IfBooleanTF{#1}
		{ 
			\IfNoValueTF{#5}
			{\bbrakket*{#2}{} \IfBooleanTF{#4}{*}{}}
			{\bbrakket*{#2}{#5}}
		}
		{
			\IfBooleanTF{#4}
			{ 
				\IfNoValueTF{#5}
				{\bbrakket{#2}{} *}
				{\bbrakket*{#2}{#5}}
			}
			{\bbrakket{#2}{\IfNoValueTF{#5}{}{#5}}} 
		}
	}
	{ 
		\IfBooleanTF{#1}
		{\vphantom{#2}\left\llangle\smash{#2}\right\rvert}
		{\left\llangle{#2}\right\rvert}
		\IfBooleanTF{#4}{*}{}
		\IfNoValueTF{#5}{}{#5}
	}
}
\DeclareDocumentCommand\kket{ s m }
{ 
	\IfBooleanTF{#1}
	{\vphantom{#2}\left\lvert\smash{#2}\right\rrangle} 
	{\left\lvert{#2}\right\rrangle} 
}

\DeclareDocumentCommand\iinnerpproduct{ s m g }
{ 
	\IfBooleanTF{#1}
	{ 
		\IfNoValueTF{#3}
		{\vphantom{#2}\left\llangle\smash{#2}\middle\vert\smash{#2}\right\rrangle}
		{\vphantom{#2#3}\left\llangle\smash{#2}\middle\vert\smash{#3}\right\rrangle}
	}
	{ 
		\IfNoValueTF{#3}
		{\left\llangle{#2}\middle\vert{#2}\right\rrangle}
		{\left\llangle{#2}\middle\vert{#3}\right\rrangle}
	}
}
\DeclareDocumentCommand\bbrakket{}{\iinnerpproduct} 
\DeclareDocumentCommand\iip{}{\iinnerpproduct} 
	
\DeclareDocumentCommand\oouterpproduct{ s m g }
{ 
	\IfBooleanTF{#1}
	{ 
		\IfNoValueTF{#3}
		{\vphantom{#2}\left\lvert\smash{#2}\middle\rrangle\!\middle\llangle\smash{#2}\right\rvert}
		{\vphantom{#2#3}\left\lvert\smash{#2}\middle\rrangle\!\middle\llangle\smash{#3}\right\rvert}
	}
	{ 
		\IfNoValueTF{#3}
		{\left\lvert{#2}\middle\rrangle\!\middle\llangle{#2}\right\rvert}
		{\left\lvert{#2}\middle\rrangle\!\middle\llangle{#3}\right\rvert}
	}
}

\DeclareDocumentCommand\eexpectationvalue{ s s m g }
{ 
	\IfNoValueTF{#4}
	{
		\IfBooleanTF{#1}
		{\vphantom{#3}\left\llangle\smash{#3}\right\rrangle} 
		{\left\llangle{#3}\right\rrangle} 
	}
	{
		\IfBooleanTF{#1}
		{
			\IfBooleanTF{#2}
			{\left\llangle{#4}\middle\vert{#3}\middle\vert{#4}\right\rrangle} 
			{\vphantom{#3#4}\left\llangle\smash{#4}\middle\vert\smash{#3}\middle\vert\smash{#4}\right\rrangle} 
		}
		{\vphantom{#3}\left\llangle{#4}\middle\vert\smash{#3}\middle\vert{#4}\right\rrangle} 
	}
}
\DeclareDocumentCommand\eev{}{\eexpectationvalue} 

\DeclareDocumentCommand\mmatrixelement{ s s m m m }
{ 
	\IfBooleanTF{#1}
	{
		\IfBooleanTF{#2}
		{\left\llangle{#3}\middle\vert{#4}\middle\vert{#5}\right\rrangle} 
		{\vphantom{#3#4#5}\left\llangle\smash{#3}\middle\vert\smash{#4}\middle\vert\smash{#5}\right\rrangle} 
	}
	{\vphantom{#4}\left\llangle{#3}\middle\vert\smash{#4}\middle\vert{#5}\right\rrangle} 
}
\DeclareDocumentCommand\mmel{}{\mmatrixelement} 

\DeclareMathOperator*{\crit}{crit}
\begin{document}

\begin{abstract}
	At the moment, the most efficient method to compute the state of a periodically driven quantum system is using Floquet theory and the Floquet eigenbasis. The wide application of this basis set method is limited by: a lack of unique ordering of the Floquet eigenfunctions, an ambiguity in their definition at resonance, and an instability against infinitesimal perturbation at resonance. We address these problems by redefining the eigenbasis using a revised definition of the average energy as a quantum number. As a result of this redefinition, we also obtain a Floquet-Ritz variational principle, and justify the truncation of the Hilbert space.
\end{abstract}

\title{Defining a well-ordered Floquet basis by the average energy}

\author{Cristian M. \surname{Le}}
\email{cristian.le@phys.s.u-tokyo.ac.jp}
\author{Ryosuke \surname{Akashi}}
\affiliation{Department of Physics, University of Tokyo, Hongo, Tokyo, 113-0033, Japan}
\author{Shinji \surname{Tsuneyuki}}
\affiliation{Department of Physics, University of Tokyo, Hongo, Tokyo, 113-0033, Japan}
\affiliation{ISSP, University of Tokyo, Kashiwa, Tokyo, 277-8581 Japan}

\date{\today}
\maketitle
\section{Introduction}
	The works of Shirley and Sambe \cite{Shirley_1965, Sambe_1973} introducing Floquet theory to quantum mechanics, have enabled the efficient calculation of the time-periodic Schr\"{o}dinger equation and kicked off renewed interest in periodically driven quantum systems (Floquet systems). The Floquet method has since become common place when studying model systems with periodic driving \cite{[][{, and references therein}]Eckardt_2017,Holthaus_2015,[][{, and references therein}]Oka_2019}. Applications beyond model systems, and towards realistic systems derived from first-principles, are still limited by a lack of a proper definition of the ground state and a variational method which can derive it efficiently.
	
\begin{figure}
		\includegraphics[width=8.6cm]{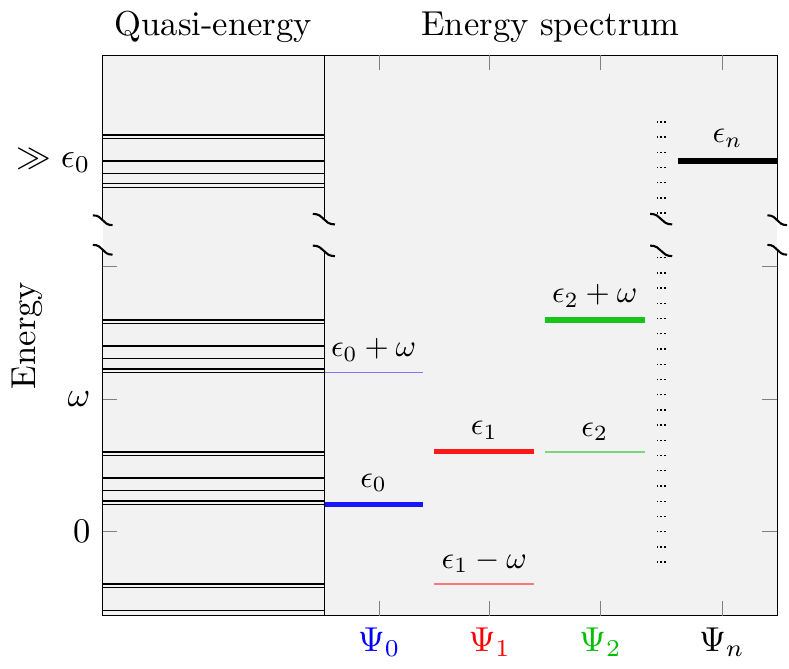}
		\caption{Quasi-energies and full energy spectra (\cref{Eq: Floq Eq,Eq: Energy spectrum}) of a typical Floquet system. The energy spectrum amplitude of each state is depicted by varying thickness and opacity.\label{Fig: Typical Floquet}}
	\end{figure}

	We condense the common problems of the Floquet method in the left side of \cref{Fig: Typical Floquet}, along with an often disregarded, yet intuitive way to solve them in the right side of the figure. Just like the energy levels of static systems, Floquet systems  are characterized by the quasi-energies $\epsilon_n$ and their corresponding Floquet eigenfunctions $u_n(t)$.
	\begin{equation}
		\qty\big[\hat{H}(t)-i\partial_t]\ket{u_n(t)}=\epsilon_n\ket{u_n(t)}\label{Eq: Floq Eq},
	\end{equation}
	where $H(t)=H(t+T)$ denotes the time-periodic Hamiltonian of the system we wish to study. Complications arise from the appearance of countably infinite equivalent solutions to \cref{Eq: Floq Eq}, with quasi-energies offset by a multiple of the driving frequency $\omega=2\pi/T$, but representing the same physical state, i.e. the same solution of the original Schr\"{o}dinger equation. Thus it is impossible to label the eigenstates and define a ground state by the quasi-energy alone, as it is made apparent in the left side of \cref{Fig: Typical Floquet}.

	On the other hand we find hints of ordering when we consider the eigenstate's energy spectrum, as defined by:
	\begin{gather}
	 	P^{\infty}_n(E)=\abs{\frac{1}{2\pi}\int_{-\infty}^{+\infty}e^{i(E-\epsilon_n)t}\ket{u_n(t)}\dd{t}}^2.\label{Eq: Energy spectrum}
	\end{gather}
	We show this in the right side of \cref{Fig: Typical Floquet} by varying the width and opacity of the lines. The eigenstates in \cref{Fig: Typical Floquet} are chosen to depict three common features present in a typical Floquet system. Firstly we can see that a ground state can be uniquely defined (state $\Psi_0$), given a robust enough criterion, e.g. ordering by the average of the spectrum. Secondly the energy spectrum is plagued with near resonant states which creates an ambiguity in how to define the eigenstates. This is similar to the degeneracy problem of static systems, e.g. quantization axis of the Hydrogen's $\mathrm{2p^1}$ states. In the Floquet system here, resonant states $\Psi_1$ and $\Psi_2$ are not uniquely defined from the solutions of \cref{Eq: Floq Eq}. This ambiguity can be resolved by using additional quantum numbers, and the ones based on the energy spectrum are the most intuitive. Lastly there are numerous states $\Psi_n$ with relatively similar quasi-energies, but well separated energy spectra compared to the low-lying/active states.  Such states should not affect the physical system significantly and we can truncate the Hilbert space and significantly simplify the calculations needed.
	
	In this paper we propose the average energy to be the central parameter to label and define the Floquet eigenstates. We show that this definition has a variational derivation, is stable against infinitesimal perturbations and allows for the truncation of the Hilbert space. Before diving into this investigation, we present a short history of the problem of defining a Floquet ground state in \cref{Sec: History}, where the reader can find various alternatives that are currently used. We summarize the potentially ambiguous notations in \cref{Sec: Notation} for the reader to refer back to if the equations become unclear. At the beginning of \cref{Sec: Floquet systems} we present the basic derivation of the Floquet method with special emphasis on the problems concerning the eigenstates definition. The infinite time average energy and its corresponding eigenstates are defined in \cref{Sec: Exact average energy}. For practical applications we derive the observed average energy in \cref{Sec: Approximate average energy}, with the reasoning explained at the start of \cref{Sec: Avoided crossing}. Using that definition we derive an equivalent Ritz variation method in \cref{Sec: Floquet-Ritz}. We use the two-level system with a circularly polarized driving as the minimal example system and exemplify the prior discussions in \cref{Sec: Example Two-level,Sec: Example perturbed two-level}.

\subsection{History of the problem\label{Sec: History}}
	The Floquet method has been thoroughly studied for decades and the problems of labeling the eigenstates and defining a ground state have been known from the beginning \cite{Shirley_1965, Sambe_1973, Kohn_2001}. So far there have been three ways of solving this problem, by labeling the eigenstates using: a Floquet-Brillouin zone in quasi-energy space \cite{Sambe_1973,Holthaus_2015}, an adiabatic continuation \cite{Young_1970, Hone_1997, Weinberg_2017}, or a perturbation method \cite{Sambe_1973, Mikami_2016, Takegoshi_2015, Rodriguez-Vega_2018}. These methods are commonly applied to finite model systems where in principle the full Hilbert space is accessible, and there is minimal need to label the eigenstates. On the other hand, these methods fail when applied to realistic systems, where the Hilbert space has infinite degrees of freedom. The incompatibilities in the realistic system boil down to the resonance condition of infinitely many eigenstates and their avoided crossing problem \cite{Hone_1997,Hone_2009}. While there have been various adiabatic methods proposed to overcome this problem \cite{Hone_1997, Weinberg_2017, Fleischer_2005}, there is still a need for eigenstate labeling methods which are efficient, do not rely on an adiabatic continuation and can be applied to realistic systems.
	
	Another common approach is to circumvent the need to identify any particular eigenstate and extend the system to an open quantum system \cite{Kohn_2001, Iadecola_2015, Hartmann_2017, Diermann_2019}. This is analogous to changing the problem from calculating the ground state to calculating the thermal equilibrium, which in this case is commonly referred to as the Floquet steady-state. This is often the most achievable and physically relevant state. Since the steady-state does not generally have a Gibbs distribution \cite{Shirai_2015, Shirai_2016, Kohn_2001, Langemeyer_2014, Iadecola_2015, Liu_2015a}, at the low temperature limit the system does not generally reduce to a pure ground state, which can diminish the significance of labeling the eigenstates and defining a ground state. However, even here, the eigenstate labeling problem is not fully circumvented, since there is still no criterion for truncating the Hilbert space. Thus the computation cost become unfeasible in realistic systems.
	
	To give an example of the problems coming from the lack of unique labeling, we look at the attempts of formulating Floquet \textit{ab-initio} methods \cite{Deb_1982, Roos_1984}. These methods were constructed to calculate a poorly defined ground state, for which the Ritz variational principle is assumed. One of these methods, the Floquet Density Functional Theory \cite{Deb_1982}, was shown to be fundamentally flawed due to the ambiguous definition of the ground state \cite{Maitra_2002, Maitra_2007, Kapoor_2013}, an argument that can be extended to the other variational based \textit{ab-initio} derivations as well. One method that remains valid is the Floquet Density Matrix Renormalization Group \cite{Zhang_2017} and similar iterative methods. But even there, the solution and its convergence is dependent on the selection criterion used at each iteration. A starting point for reformulating these \textit{ab-initio} methods is to uniquely redefine the ground state of a general Floquet system, particularly one that is applicable to realistic Hamiltonians.
	
	The potential usage of an average energy to label the eigenstates has been considered before, either indirectly \cite{Hone_2009} for selecting relevant adiabatic states, or directly \cite{Ketzmerick_2010} for deriving an effective Gibbs distribution. It should be pointed that the average energy definition there differs from the one we propose in this paper. This usage is often disregarded since other observables offer a better agreement, e.g. regular energy \cite{Ketzmerick_2010}. However this does not rule out the potential usage of the average energy to approximate and select the significant eigenstates for describing the steady state solution or truncating the Hilbert space. Using the average energy definitions from \cite{Ketzmerick_2010, Fainshtein_1978} can still be problematic as we will explain in more detail later on. To resolve these problems we derive a more robust definition of the average energy.
	
\subsection{Notations\label{Sec: Notation}}
	In this paper we use the following notations in order to keep the equations more compact. The details are clarified throughout the paper as it becomes relevant. Readers can skip this section and come back if any notations appear unclear. These notations are not in any particular order, so we just present them below as is:
	\begin{gather}
		\overset{\text{Floquet Hamiltonian:}}{\hat{H}^F(t)=\hat{H}(t)-i\partial_t},\\
		\overset{\text{Fourier decomposition:}}{u^{(k)}=\frac{1}{T}\int_{0}^{T}e^{ik\omega t}u(t)\dd{t}},\\
		\overset{\text{Average expectation:}}{\eev{\hat{\mathcal{O}}}{\Psi}_\mathcal{T}=\frac{1}{\mathcal{T}}\int_{0}^{\mathcal{T}}\ev{\hat{\mathcal{O}}(t)}{\Psi(t)}\dd{t}},\\
		\overset{\text{Energy difference}}{\omega_{mnl}=\epsilon_n-\epsilon_m+l\omega}.
	\end{gather}
	The time-dependence will often be dropped, unless explicitly needed for clarity. In general we reserve the superscripts to indicate different variations of the functionals, eigenstates, etc.. These can often be mixed with each other. If the superscript is missing any definition is applicable and/or the perturbed/untruncated definition is used, depending on the context.
	\begin{align}
		\bar{E}^\infty\quad&\text{\textbf{Infinite time} average energy},\\
		\bar{E}^\mathcal{T}\quad&\text{\textbf{Observed} average energy},\\
		\bar{E}^0\quad&\text{\textbf{Unperturbed} average energy},\\
		\bar{E}^i\quad&\text{\textbf{Truncated} average energy at step }i.
	\end{align}
	
	Subscripts are generally reserved to label the eigenstates, eigenvalues, etc.. We reserve the following subscript (or superscript in the case of truncation) notations:
	\begin{align}
		a,b,c,\ldots\quad&\text{Resonant states},\\
		m,n,o,\ldots\quad&\text{Ordered states},\\
		i,j,k,\ldots\quad&\text{Truncation step}.
	\end{align}
	
	$\mathcal{U}$ is reserved for the transformation matrix from an eigenstate representation to an arbitrary basis, and is constructed from the vector columns of the Floquet eigenfunction
	\begin{equation}
		\overset{\text{Transformation matrix}}{\mathcal{U}(t)=\mqty[c_{a0}(t)&c_{a1}(t)&\ldots\\
		c_{b0}(t)&c_{b1}(t)&\ldots\\
		\vdots&\vdots&\ddots]}.\label{Eq: Transf matrix}
	\end{equation}
	
	Hilbert spaces are denoted like $\mathds{H}$, and the equivalent eigenspace specific to each Hamiltonian $H$ is $\mathcal{E}_{H}$.
	\begin{equation}
		\overset{\text{Eigenspace:}}{\mathcal{E}_{H}=\qty{E_n,\Psi_n\Big\vert\; \hat{H}\ket{\Psi_n}=E_n\ket{\Psi_n};\;\ip{\Psi_m}{\Psi_n}=\delta_{mn}}}.
	\end{equation}
	
	The Floquet interaction picture with respect to interaction $v$ ($H=H^0+v$) is denoted by the subscript $_I$.
	\begin{gather}
		\overset{\text{Non-interacting propagator:}}{\begin{aligned}
			\hat{U}^0(t)=&\hat{U}^0(t,0)=e^{-i\int_{0}^{t}\hat{H}^0(\tau)\dd{\tau}},\\
			=&\sum_{n}e^{-i\epsilon^0_nt}\op*{u^0_n(t)}{u^0_n(0)}
		\end{aligned}},\\
		\overset{\text{Interaction picture propogator}}{i\partial_t\ket{u_I(t)}=\hat{v}_I(t)\ket{u_I(t)}},\\
		\overset{\text{Interaction picture operator}}{\hat{\mathcal{O}}_I(t)=\hat{U}^{0\dagger}(t)\hat{\mathcal{O}}(t)\hat{U}^0(t)}.
	\end{gather}
	
	Finally we use the usual $\delta$ notation to indicate functional variation in arbitrary direction, derivative, etc., and we add a constrained variation notation. The notation is generally omitted if we consider the full Hilbert space.
	\begin{equation}
		\overset{\text{Constrained variation:}}{\eval{\delta\mathcal{O}[u]}_{\mathds{H}}=\qty{\int\fdv{\mathcal{O}}{u}\delta u\Big\vert\delta u\in\mathds{H}}}.
	\end{equation}
	
\section{Exact Floquet systems\label{Sec: Floquet systems}}
	We start off with a thorough examination of the Floquet theory in the closed quantum systems. Consider the Schr\"{o}dinger equation of a quantum system described by the time-periodic Hamiltonian $H(t+T)=H(t)$, with a corresponding driving frequency $\omega=2\pi/T$
	\begin{equation}
		i\partial_t\ket{\Psi(t)}=\hat{H}(t)\ket{\Psi(t)}. \label{Eq: Schro}
	\end{equation}	
	Using Floquet theory we can solve this equation as an eigenproblem resembling the solution of a static Hamiltonian
	\begin{equation}
		\qty[\hat{H}(t)-i\partial_t]\ket{u_n(t)}=\epsilon_n\ket{u_n(t)}\tag{\ref{Eq: Floq Eq}}.
	\end{equation}	

	From these eigenfunctions we can derive the propagator and the wavefunction in \cref{Eq: Schro} at arbitrary times. This is analogous to the time-dependent wavefuntion solution of a static Hamiltonian.
	\begin{gather}
		\hat{U}(t)=\hat{U}(t,0)=\sum_{n}e^{-i\epsilon_nt}\op{u_n(t)}{u_n(0)},\label{Eq: Floq Propagator}\\
		\ket{\Psi(t)}=\sum_{n}C_n e^{-i\epsilon_nt}\ket{u_n(t)}, \label{Eq: Floq Phys states}\\
		C_n=\ip{u_n(0)}{\Psi(0)}.
	\end{gather}	
	Here $C_n$ is the usual overlap at a known time point $t=0$, and will be used throughout this paper. In order to preserve the norm of the wavefunction, the summations in \cref{Eq: Floq Propagator,Eq: Floq Phys states} and further on, are limited to different Floquet eigenstates, which will be clarified shortly. It is important to note that the quantum system and all of its observables are strictly determined by the wavefunction $\Psi(t)$, and not by the solutions of the Floquet Hamiltonian $H^F$ (\cref{Eq: Floq Eq})
	\begin{align}
		O(t)=&\ev{\hat{O}(t)}{\Psi(t)}\\
		=&\ev{\hat{U}^\dagger(t)\hat{O}(t)\hat{U}(t)}{\Psi(0)},\label{Eq: Floq Observables}
	\end{align}
	
	Unfortunately we cannot directly calculate this wavefunction $\Psi(t)$ or propagator $U(t)$ efficiently. Instead we calculate them indirectly through the eigenpair of quasi-energy and Floquet eigenfunction $\qty(\epsilon_n;u_n(t))$ as in \cref{Eq: Floq Propagator,Eq: Floq Phys states}. Although these eigensolutions do not have a direct physical meaning, they are computationally accessible, and sufficient to describe arbitrary states. Deriving the Floquet eigenstates is the only reliable and efficient way of describing the system and it is thus the main objective of this paper
	
	Early on, Sambe has formalized the Floquet system's extended Hilbert space $\mathds{HT}\ni u(t)$, upon which the Floquet Hamiltonian operates and defines the eigenspace $\mathcal{E}_{H}$ \cite{Sambe_1973}. This extended Hilbert space is the tensor product of the original Hilbert space $\mathds{H}$ on which $H(t)$ operates at any given time, and the countably infinite Fourier space $\mathds{T}$ which guarantee the periodicity of the Floquet functions, in accord with Floquet theory \cite{Floquet_1883}. In this extended Hilbert space, the Floquet Schr\"{o}dinger equation has a systematic block matrix form:
	\begin{gather}
		\hat{H}^F(t)=\mqty[\ddots&\vdots\\
		\ldots&\hat{H}^{(0)}+\omega\mathds{1}&\hat{H}^{(1)}&\hat{H}^{(2)}&\\
		&\hat{H}^{(-1)}&\hat{H}^{(0)}&\hat{H}^{(1)}&\\
		&\hat{H}^{(-2)}&\hat{H}^{(-1)}&\hat{H}^{(0)}-\omega\mathds{1}&\ldots\\
		&&&\vdots&\ddots],\label{Eq: Floq Ham Matrix}\\
		\ket{u(t)}=\mqty(\ldots&\ket*{u^{(-1)}}e^{i\omega t}&\ket*{u^{(0)}}&\ket*{u^{(1)}}e^{-i\omega t}&\ldots)^T.
	\end{gather}	
	The Floquet eigenfunctions diagonalize this Hamiltonian, and form an orthonormal complete basis set spanning $\mathds{HT}$.
	\begin{gather}
		\mmel{u_m}{\hat{H}^F}{u_n}_T=\epsilon_n\delta_{mn}\label{Eq: Floq basis space0},\\
		\llangle u_m\vert u_n\rrangle_T=\delta_{mn}\label{Eq: Floq basis space1}.
	\end{gather}
	
	Exactly diagonalizing the infinite matrix in \cref{Eq: Floq Ham Matrix} is usually a difficult problem, even when the Hamiltonian's Hilbert space $\mathds{H}$ is finite and discrete. Exact solutions are limited to some of the well-known model systems: Two-level system \cite{Schmidt_2019, Xie_2017}, harmonic oscillator \cite{Breuer_1989, Hanggi_1998}, free electron \cite{Madsen_2005, Joachain_2011}. We can expand the solvable systems with the use of perturbation methods, e.g. the weak interaction \cite{Sambe_1973}, high-frequency \cite{Mikami_2016, Casas_2001, Eckardt_2015}, low frequency \cite{Rodriguez-Vega_2018}, and continued fraction expansions \cite{Hanggi_1998, Giovannini_2019}.
	
	Sambe also showed that a variational principle on the quasi-energy is possible, with the stationary points corresponding to the Floquet eigenfunctions.
	\begin{gather}
		\overset{\text{Variational principle:}}{\var{\epsilon[u]}=0\Rightarrow u(t)\in\qty{u_n(t)}}\label{Eq: quasi variation},\\
		\epsilon[u]=\eev{\hat{H}^F}{u}_T.
	\end{gather}
	This variational principle offers an alternative method of deriving the Floquet eigenfunctions, which is more computationally efficient and necessary for many theoretical formulations. From here on, we will assume that the exact eigensolutions $\qty{\epsilon_n,u_n(t)}$ are known.
	
	The infinite dimension of the Floquet Hamiltonian in Fourier space $\mathds{T}$ suggest that there are a countable infinite number of eigenpairs $\qty{\epsilon'_n;u'_n(t)}$. These infinite solutions can be grouped into subsets, with the solutions related to each other by harmonic shifts (\cref{Eq: Alt basis}), and describing the same wavefunction $\Psi_n(t)$ (\cref{Eq: Full Floq states}).
	\begin{gather}
		\ket*{u_n^{\prime(k)}}=\ket*{u_n^{(k+l)}};\quad\epsilon'_n=\epsilon_n+l\omega \qquad\forall\;l\in\mathds{Z} \label{Eq: Alt basis},\\
		\ket{\Psi'_n(t)}=\ket{\Psi_n(t)}=e^{-i\epsilon_nt}\ket{u_n(t)}\label{Eq: Full Floq states},\\
		\ip{\Psi_m(t)}{\Psi_n(t)}=\delta_{mn}\quad\forall\; t.
	\end{gather}
	We refer to the wavefunction $\Psi_n(t)$ as the physical Floquet eigenstate, to distinguish it from the ambiguous Floquet eigenfunction $u_n(t)$. The physical Floquet eigenstates form a complete basis set and fully describe the propagator in \cref{Eq: Floq Phys states}. These basis sets span the Hamiltonian's Hilbert space $\mathds{H}$ instead of the extended one $\mathds{HT}$.
	
	Because of this ambiguity, only a subset of the eigenfunctions $u_n(t)$ are needed in summations like \cref{Eq: Floq Propagator,Eq: Floq Phys states} in order to span the Hilbert space $\mathds{H}$ and describe the propagation of arbitrary wavefunctions. The choice of eigenfunctions is arbitrary, as long as the Hilbert space $\mathds{H}$ is fully spanned, or equivalently there are no pairs related to each other by a harmonic shift (\cref{Eq: Alt basis}). We refer to these choices in eigenbasis as a choice of quasi-energy shift. Any observable, interaction, etc. are independent of this choice, except for the quasi-energy $\epsilon$ which is not a true observable according to \cref{Eq: Floq Observables}. From here on we assume that an arbitrary choice of quasi-energy shift is performed and labels $m,n$ describe distinct physical Floquet eigenstates.
	
	Another important consequence of the expanded Hilbert space, is that the Floquet eigenfunctions are not uniquely defined within the subspace of resonant eigenfunctions. That is to say, we can rotate the basis $\qty{u_a(t)}$ within the resonant subspace $\epsilon_a=\epsilon_b=\epsilon$, and obtain another orthonormalized basis$\iip*{u'_a}{u'_b}_T=\delta_{ab}$, which is an equally valid eigenbasis of the Floquet Hamiltonian $H^F$.
	\begin{align}
		\ket{u'_a(t)}=&\sum_b C'_{ba}\ket{u_b(t)} &\forall&\epsilon_b=\epsilon\label{Eq: Floq resonant mix},\\
		\delta \epsilon[u'_a]=&\delta\epsilon[u_b]=0 &\forall& u'_a(t)\in\mathds{HT}_\epsilon,\\
		\eev{\hat{H}^F}{u'_a}_T=&\eev{\hat{H}^F}{u_b}_T=\epsilon &\forall& u'_a(t)\in\mathds{HT}_\epsilon\label{Eq: Floq resonant eigen}.
	\end{align}
	In this paper we reserve the labels $a,b$ to describe resonant Floquet eigenfunctions, and we assume that we have selected the subset of eigenfunctions so that the resonant eigenfunctions have the same quasi-energy $\epsilon_a=\epsilon_b=\epsilon$. For each resonant set with quasi-energy $\epsilon$, $\mathds{HT}_\epsilon$ denotes the resonant Hilbert subspace to which it belongs.
	
	This ambiguity problem is analogous to the degeneracy problem of static systems, where for example in the Hydrogen $\mathrm{2p}^1$ degenerate space, we can arbitrarily choose the axis of quantization, and any choice gives valid Hamiltonian eigenbasis. However there are two caveats in the Floquet system compared to the static case. Firstly, the number or density of resonant Floquet eigenfunctions is practically infinite in realistic systems, e.g. the Volkov states \cite{Joachain_2011, Kidd_2018}, while in general the static system is finite, with the exception of flat band systems which are still resolvable. Secondly, the energy spectrum as defined in \cref{Eq: Energy spectrum} differs between different resonant Floquet eigenfunctions (\cref{Fig: Resonance1}), and is dependent on our choice of eigenbasis.
	\begin{equation}
		P'_a(E)\neq P_b(E)\qquad \forall\;a\neq b.
	\end{equation}
	A consequence of this energy spectrum difference is that, if we include thermodynamic effects, resonant eigenstates interact differently and are not equally distributed at equilibrium \footnote{Part of the occupation difference comes from the bath spectrum dependence of the Floquet steady-state \cite{Iadecola_2015, Kohn_2001, Langemeyer_2014}, but even accounting for it (e.g. a flat system-bath interaction) the difference persists}. In the extreme case of highly separated energy spectrum, we can find a resonant eigenbasis where only a few states are occupied in the steady-state, so that we can truncate the Hilbert space without affecting the description of the steady state. If resonant eigenfunctions share the same energy spectrum, these are indeed degenerate.
	
	In the case of a static system, we can use an additional quantum number or an adiabatic continuation to label the eigenstates at and around the degeneracy point. In limited cases \cite{Hone_1997}, we can also use the adiabatic method to define the Floquet eigenfunctions around the resonance (e.g. \cref{Fig: TLS}). On the other hand, defining the eigenstates using an additional quantum number would be more efficient, so we aim to find an appropriate parameter that would result in reasonable and/or truncatable eigenbasis (e.g. \cref{Fig: Typical Floquet}). The most natural choice is one based on the energy spectrum, and the first possible choice is the average energy.
	
	\subsection{Infinite time average energy\label{Sec: Exact average energy}}
	We redefine the average energy as simply the time-averaged expectation value of the Hamiltonian of an arbitrary state/initial wavefunction up to some time $\mathcal{T}$, to be clarified later in the paper. This definition gives a proper observable (\cref{Eq: Floq Observables}), and as such, it is independent of our choice of the quasi-energy shift.
	\begin{align}
		\bar{E}^\mathcal{T}[\Psi(0)]=&\eev{\hat{H}}{\Psi}_\mathcal{T}\label{Eq: Average energy def}\\
		=&\frac{1}{\mathcal{T}}\int_{0}^{\mathcal{T}}\ev{\hat{U}^\dagger(t)\hat{H}(t)\hat{U}(t)}{\Psi(0)}\dd{t}\notag.
	\end{align}
	This definition is closely related to the energy spectrum average (\cref{Eq: Average energy def2}), and can be adapted for various theoretical formulations.
	\begin{gather}
		\bar{E}^\mathcal{T}[\Psi(0)]\approx\int_{-\infty}^{+\infty}P^\mathcal{T}[\Psi(0),E]\dd{E}\label{Eq: Average energy def2},\\
		P^\mathcal{T}[\Psi(0),E]=\abs{\frac{1}{2\pi}\int_{-\mathcal{T}}^{\mathcal{T}}e^{iEt}\hat{U}(t)\ket{\Psi(0)}\dd{t}}^2\label{Eq: Energy Spec def2}.
	\end{gather}
	
	Before defining the averaging time $\mathcal{T}$, we will address the previous definition of the average energy \cite{Fainshtein_1978, Ketzmerick_2010}. There the average energy has been defined as the average energy expectation value over a period $T$, only of a Floquet eigenfunction $u_n(t)$ (derived from \cref{Eq: Floq Eq}).
	\begin{equation}
		\bar{\epsilon}^T_n=\eev{\hat{H}}{u_n}_T=\epsilon_n+\sum_{k}k\omega\ip*{u_n^{(k)}}\label{Eq: Eigen avg energy summation}.
	\end{equation}
	This value is also independent of our choice of quasi-energy shift (\cref{Eq: Alt basis}) and thus could be used for labeling the eigenstates. An obvious flaw of this definition is that it is not defined at resonance, since the Floquet eigenstate itself is not uniquely defined there. Another problem is that, the generalization of this definition to arbitrary states (\cref{Eq: Effective average energy}) is incompatible with the variational principle, and is not defined by the Floquet eigenstate's average energies $\bar{\epsilon}_n$ (regardless of their definition). Nevertheless, we refer to this generalization as the effective average energy, to be used in later derivations.
	\begin{gather}
		\overset{\text{Effective average energy:}}{\bar{\epsilon}^T[u]=\frac{1}{T}\int_{0}^{T}\ev{\hat{H}(t)}{u(t)}\dd{t}}\label{Eq: Effective average energy},\\
		\delta\bar{\epsilon}^T[u]=0\nRightarrow u(t)\in\qty{u_n(t)},\\
		\bar{\epsilon}^T[u]\neq \abs{C_n}^2\bar{\epsilon}_n.
	\end{gather}
	This suggest that we cannot use this average energy definition to derive a Floquet eigenstate directly, and we need a different average energy definition.
	
	Coming back to the definition in \cref{Eq: Average energy def}, we can simplify the equation by using the exact propagator (\cref{Eq: Floq Propagator}) expanded in an arbitrary choice of eigenfunctions:
	\begin{align}
		\bar{E}^\mathcal{T}=&\sum_{n}\abs{C_n}^2\eev{\hat{H}}{u_n}_\mathcal{T}\notag\\
		&+\sum_{m,n}C^*_mC_n\frac{1}{\mathcal{T}}\int_{0}^{\mathcal{T}}e^{-i(\epsilon_n-\epsilon_m)t}\mel{u_m}{i\partial_t}{u_n}\dd{t}\label{Eq: Floq Avg Energy initial}.
	\end{align}
	
	In this section we choose the averaging time $\mathcal{T}$ to be the limit at infinity, and refer to the resulting average energy as the infinite time average energy, and equivalently the eigenstates derived from it.
	\begin{equation}
		\bar{E}^\infty[\Psi(0)]=\lim_{\mathcal{T}\to\infty}\bar{E}^\mathcal{T}[\Psi(0)]\label{Eq: Floq Infiinite Avg Energy def}.
	\end{equation}
	
	For now we assume there are no resonant eigenfunctions $(\text{mod}(\epsilon_n-\epsilon_m,\omega)\neq0)$. The infinite time average energy functional trivially simplifies to the weighted sum of the eigenstate average energies
	\begin{equation}
		\bar{E}^\infty[\Psi(0)]=\sum_{n}\abs{C_n}^2\bar{\epsilon}^{\infty}_n\qquad if\;\nexists\;\epsilon_n=\epsilon_m\label{Eq: Floq Avg Energy decomp}.
	\end{equation}
	Where due to non-resonance the eigenstate average energies are unambiguously defined as either the effective or infinite time value $\bar{\epsilon}^T_n=\bar{\epsilon}^\infty_n=\bar{\epsilon}_n$. Immediately we see the benefit of this definition that we can derive the eigenstates from the variation of the functional form. The variational principle in this case is over the initial wavefunction $\Psi(0)$, as opposed to the previous variations over the Floquet functions $u(t)$.
	\begin{equation}
		\delta\bar{E}^\infty[\Psi(0)]=0\Rightarrow\Psi(0)\in\qty{u_n(0)};\;\bar{E}^\infty\in\qty{\bar{\epsilon}^\infty_n}.
	\end{equation}
	
	In theory this variation is sufficient to define and find all eigenpairs $\qty{\bar{\epsilon}^\infty_n,\Psi_n(0)=u_n(0)}$, which form a well ordered eigenspace $\mathcal{E}^{\infty0}_H$ with $\bar{\epsilon}^\infty_0\leq\bar{\epsilon}^\infty_1\leq\ldots$.
	\begin{equation}
		\mathcal{E}_{H}^{\infty0}=\qty{\bar{\epsilon}^\infty_n,\Psi_n(0)\Big\vert\delta\bar{E}^\infty[\Psi]=0}\qquad if\,\nexists\,\epsilon_n=\epsilon_m\label{Eq: Eigenspace Initial WF}.
	\end{equation}
	The superscript in $\mathcal{E}^0_{H}$ indicates that only the solutions at $t=0$ are calculated and the Hilbert space is limited to $\mathds{H}$, not to be confused with the notation $\mathcal{E}_{H^0}$. From these we can also derive the time-dependent Floquet functions $u_n(t)$ and quasi-energies $\epsilon_n$ from the Schr\"{o}dinger equation, and subsequently the propagator and any observable of arbitrary states at arbitrary times.
	
	In practice however, the exact propagator $U(t)$ in \cref{Eq: Average energy def} is inaccessible, and the variation with respect to the initial wavefucntion $\Psi(0)$ is computationally unfeasible. Fortunately as long as there are no resonance conditions, we can use the quasi-energy variation principle (\cref{Eq: quasi variation}) to derive the Floquet eigenstates and the effective average energy (\cref{Eq: Effective average energy}) which coincide with the infinite time average energy for the non-resonant Floquet eigenstates.
	\begin{equation}
		\bar{\epsilon}^T[u_n(t)]=\bar{E}^\infty[\Psi_n(0)]\in\qty{\bar{\epsilon}^T[u]\Big\vert\delta\epsilon[u]=0}.
	\end{equation}
	
	Thus we can define a well ordered eigentriplet $\qty(\epsilon_n,\bar{\epsilon}^\infty_n,u_n(t))$ with $\bar{\epsilon}^\infty_0\leq\bar{\epsilon}^\infty_1\leq\ldots$, up to a trivial quasi-energy shift. The correspondent eigenspace $\mathcal{E}^\infty_{H}$ spans the extended Hilbert space $\mathds{HT}$ (having included equivalent solutions \cref{Eq: Alt basis}), and the propagator is directly obtained from \cref{Eq: Floq Propagator}.
	\begin{equation}
		\mathcal{E}^\infty_{H}=\qty{\epsilon_n,\bar{\epsilon}^\infty_n,u_n(t)\Big\vert\delta\epsilon[u]=0}\qquad if\,\nexists\,\epsilon_n=\epsilon_m.
	\end{equation}
	
	From these eigentriplets we can uniquely define the ground state $\qty(\epsilon_0,\bar{\epsilon}^\infty_0,u_0(t))$, up to a trivial quasi-energy shift. This ground state can be derived variationaly through the initial wavefunction $\Psi(0)$, Floquet function $u(t)$, or using a Lagrange minimization method.
	\begin{gather}
		\bar{\epsilon}^\infty_0=\min_{\Psi(0)}\bar{E}^\infty[\Psi(0)]\beginsubeqn,\\
		\bar{\epsilon}^\infty_0=\min_{u(t)}\qty{\bar{\epsilon}^T[u]\Big\vert\delta\epsilon[u]=0}\subeqn,\\
		\bar{\epsilon}^\infty_0=\min_{u(t)}\qty\Big{\bar{\epsilon}^T[u]+\lambda\fdv{\epsilon[u]}{u}+\bar{\epsilon}(\iip{u}-1)}\subeqn\label{Eq: Lagrange min method}.
	\end{gather}
	The Lagrange multiplier $\lambda$ and functional derivative $\fdv{\epsilon[u]}{u}$ are vectors spanning the Hilbert space dimensions, and guarantee the minimization in $\bar{\epsilon}^T[u]$ is taken over the Floquet eigenfunctions. The Lagrange multiplier $\bar{\epsilon}$ guarantees the normalization constraint is satisfied. Although this form is more complicated with the addition of $\lambda$, it can facilitate the derivation of other approximations, e.g. a Hartree-Fock variant, and we include it here for future reference.
	
	As for the resonant region, we cannot use the Floquet eigenfunctions defined by \cref{Eq: Floq Eq} to simplify \cref{Eq: Floq Avg Energy initial} to \cref{Eq: Floq Avg Energy decomp}, since the eigenfunctions themselves are not uniquely defined in this region. This also means that the eigenstate average energies are not well defined from \cref{Eq: Eigen avg energy summation}, and we have to redefine them here. We limit ourselves now to the resonant Hilbert subspace $\mathds{HT}_\epsilon$.
	\begin{align}
		\eev{\hat{H}^F}{u}_T=\epsilon\quad&\forall u\in\mathds{HT}_\epsilon,\\
		\eval{\delta\epsilon[u]}_{\mathds{HT}}=0\quad&\forall u\in\mathds{HT}_\epsilon.
	\end{align}
	Choosing an arbitrary orthonormal basis set in this subspace $\ev{\ip{u_a}{u_b}}_T=\delta_{ab}$, the infinite time average energy has the following form:
	\begin{equation}
		\bar{E}^\infty[\Psi]=\epsilon+\sum_{a,b}C^*_aC_b\sum_{k}k\omega\ip*{u_a^{(k)}}{u_b^{(k)}}\label{Eq: Average energy resonance2}.
	\end{equation}
	
	We point out that the effective average energy functional $\bar{\epsilon}^T$ (\cref{Eq: Effective average energy}) is equivalent to the infinite time average energy functional $\bar{E}^\infty$ (\cref{Eq: Average energy resonance2}) within the resonant Hilbert space, which is why we refer to it as effective average energy.
	\begin{align}
		\bar{E}^\infty[\Psi_a]=\bar{\epsilon}^T[u_a]\qquad&\forall u_a(t)\in\mathds{HT}_\epsilon\label{Eq: Effective Average Propert0},\\
		\eval{\delta\bar{E}^\infty[\Psi_a]}_{\mathds{HT}_\epsilon}=\eval{\delta\bar{\epsilon}^T[u_a]}_{\mathds{HT}_\epsilon}\qquad&\forall u_a(t)\in\mathds{HT}_\epsilon\label{Eq: Effective Average Propert1}.
	\end{align}
	
	\Cref{Eq: Average energy resonance2} can be represented in a matrix form, whose eigenvalues $\bar{\epsilon}^\infty_n$ are independent of the basis set. Diagonalizing this matrix we find a unique eigenbasis, from which we uniquely define the eigentriplet $\qty(\epsilon,\bar{\epsilon}^\infty_n,u_n(t))$.
	\begin{gather}
		\bar{E}^\infty=\epsilon+\mathbf{C}^\dagger\mathbf{M}\mathbf{C}=\sum_{n}\abs{C_n}^2\bar{\epsilon}^\infty_n\label{Eq: Diagonal inf average energy},\\
		M_{ab}=\sum_{k}k\omega\ip*{u_a^{(k)}}{u_b^{(k)}},\\
		\ket{u_n(t)}=\sum_{a}c_{an}\ket{u_a(t)},\\
		\ket{\Psi_n(0)}=\sum_{a}c_{an}\ket{u_a(0)}.
	\end{gather}
	The coefficients $c_{an}$ are the corresponding eigenvector projection onto the arbitrary basis $\qty{u_a(t)}$. We label the eigenstates here with $m,n$ as to indicate that the ambiguity coming from the resonance is resolved and the eigenstate average energies are defined from the diagonalization of \cref{Eq: Diagonal inf average energy}. Considering the relation to the energy spectrum, this diagonalization gives us the basis set with the least overlap between their energy spectra and maximally separated average energies (\cref{Fig: Resonance1}). The average energies themselves can have real and accidental degeneracies, which can be further resolved by expanding this procedure to the energy variance and other higher order moments.
	
	Combining all of the resonant and non-resonant subspaces, we can define the well-ordered eigenspace $\mathcal{E}_H$ of an arbitrary Hamiltonian $H(t)$. The theoretically sufficient eigenspace defined from \cref{Eq: Eigenspace Initial WF} remains unchanged upon the combination of the Hilbert subspaces, where now the eigenstate average energy $\bar{\epsilon}^\infty_n$ is uniquely defined by the infinite time average energy variation/diagonalization:
	\begin{gather}
		\bar{\epsilon}^\infty_n\equiv\bar{E}^\infty[\Psi_n]\in\qty{\bar{E}^\infty[\Psi]\Big\vert\delta\bar{E}^\infty[\Psi]=0}\label{Eq: Eigen Average energy def0},\\
		\mathcal{E}_{H}^{\infty0}\equiv\qty{\bar{\epsilon}^\infty_n,\Psi_n(0)\Big\vert\delta\bar{E}^\infty=0}\label{Eq: Eigenspace Initial WF2}.
	\end{gather}
	
	As for a computationally accessible form, we have to rely on the quasi-energy variation. Since the infinite time average energy and effective average energy are equivalent in the resonant space, we can derive the variational principle in two steps over quasi-energy and effective average energy functionals.
	\begin{gather}
		\epsilon_n\equiv\qty\Big{\epsilon[u]\Big\vert \eval{\delta\epsilon[u]}_{\mathds{HT}}=0}\Rightarrow\mathds{HT}_{\epsilon_n}\label{Eq: Basic variation principle}\beginsubeqn,\\
		\bar{\epsilon}^\infty_n\equiv\qty\Big{\bar{\epsilon}^T[u_a]\Big\vert \eval{\delta\bar{\epsilon}^T[u_a]}_{\mathds{HT}_{\epsilon_n}}=0}\subeqn\label{Eq: Basic variation principle3},\\
		\mathcal{E}^\infty_{H}\equiv\qty\Big{\epsilon_n,\bar{\epsilon}^\infty_n,u_n(t)\Big\vert\delta\epsilon[u]=0\rightarrow\eval{\delta\bar{\epsilon}^T[u]}_{\mathds{HT}_{\epsilon_n}}=0}\label{Eq: Eigenspace Floquet Function}.
	\end{gather}
	We have to be careful of the order of the variations so as to properly limit the variation space of the effective average energy. The variation in \cref{Eq: Basic variation principle3} is not valid in the full extended Hilbert space.
	
	Equivalently we can absorb these complications into a Lagrange multiplier, and extend the method in \cref{Eq: Lagrange min method}.
	\begin{equation}
		\bar{\epsilon}^\infty_n=\crit_{u(t)}\qty\Big{\bar{\epsilon}^T[u]+\lambda\fdv{\epsilon[u]}{u}+\bar{\epsilon}\qty\Big(\iip{u}-1)}\label{Eq: Lagrange method},
	\end{equation}
	where $\crit$ indicates the search of a critical point, not only a minimum.
	
	So far we have a unique description of the Floquet eigenbasis up to a trivial quasi-energy shift $\mathcal{E}^\infty_{H}=\qty{\epsilon_n,\bar{\epsilon}^\infty_n,u_n(t)}$. This basis set describes the propagator (\cref{Eq: Floq Propagator}) uniquely, is compatible with the variational principle, is well ordered through the average energy, and is unambiguous at resonance.
	
	\subsection{Exact two-level system example\label{Sec: Example Two-level}}
	\begin{figure*}
		\hfill\subfloat[\label{Fig: TLS}]{%
			\subfigimg[scale=1.]{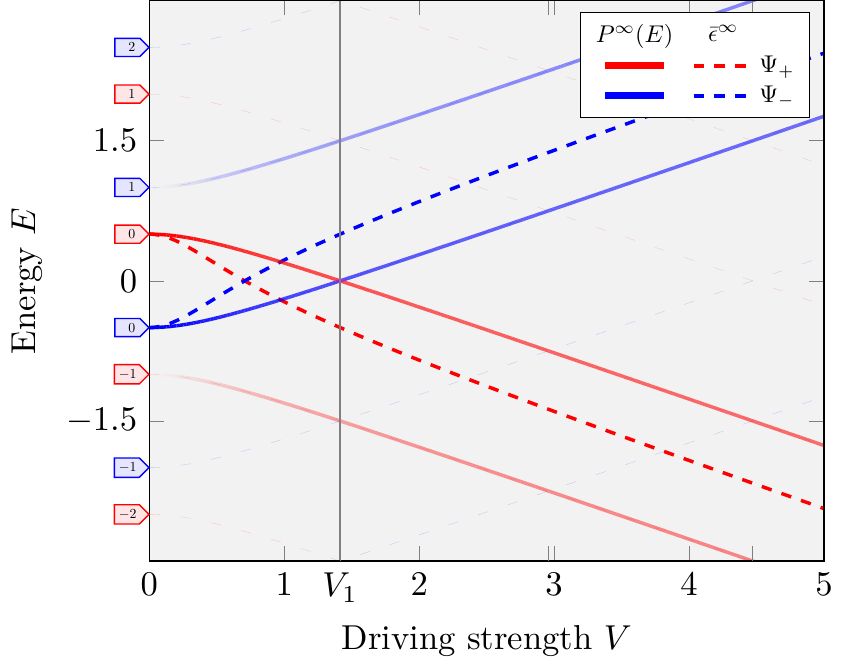}{10pt}{1.5}%
		}\hfill
		\subfloat[\label{Fig: Resonance1}]{%
			\subfigimg[scale=1.]{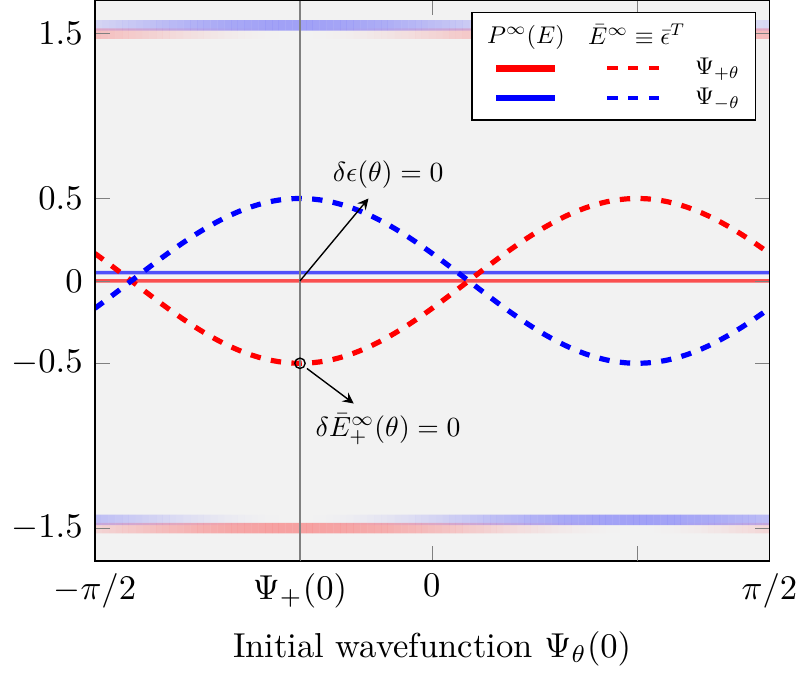}{30pt}{1.5}%
		}\hfill
		\caption{Exact energy spectrum $P^\infty(E)$ (solid line with varying intensity) and infinite time average energy $\bar{E}^\infty$ (thick dashed line) of: \protect\subref{Fig: TLS} Adiabatically connected Floquet eigenstates (\cref{Eq: Two-level eigen 0,Eq: Two-level eigen 1}); \protect\subref{Fig: Resonance1} States at the first resonance point $V=V_1$, with varying initial wavefunction $\Psi_\theta(0)$ or mixing of resonant Floquet eigenfunctions $u_\theta(t)$ (\cref{Eq: Initial WF variable definition}). The energy spectrum of the state $\Psi_{-\theta}(0)$ in \protect\subref{Fig: Resonance1} has been artificially shifted upwards for clarity. The vertical lines in both plots correspond to the eigenstate solution at the the first resonance point $V_1$, where both plots correspond to each other.\label{Fig: TLS0}}
	\end{figure*}
	The simplest toy model that shows the properties discussed above is the two-level system driven by a circularly polarized interaction. The exact solutions can be found in various textbooks \cite{Joachain_2011}, derived in a number of ways. We choose the adiabatically connected solution with varying driving strength.
	\begin{gather}
		\hat{H}(t)=\mqty[\frac{\omega_0}{2}&\frac{V}{2}e^{-i\omega t}\\\frac{V}{2}e^{+i\omega t}&-\frac{\omega_0}{2}]\label{Eq: Two-level Hamiltonian},\\
		\ket{\Psi_+(t)}=e^{-i\epsilon_+t}\mqty(\phantom{+}\sqrt{\frac{\Omega+\delta}{2\Omega}}\phantom{e^{-i\omega t}}\\
		-\sqrt{\frac{\Omega-\delta}{2\Omega}}e^{+i\omega t})\beginsubeqn\label{Eq: Two-level eigen 0},\\
		\ket{\Psi_-(t)}=e^{-i\epsilon_-t}\mqty(\phantom{+}\sqrt{\frac{\Omega-\delta}{2\Omega}}e^{-i\omega t}\\
		\phantom{+}\sqrt{\frac{\Omega+\delta}{2\Omega}}\phantom{e^{+i\omega t}})\subeqn\label{Eq: Two-level eigen 1},\\
		\epsilon_\pm=\mp\frac{\Omega-\omega}{2},\\
		\Omega=\sqrt{V^2+\delta^2}\quad\delta=\omega-\omega_0.
	\end{gather}
	
	Here the parameters $\omega_0,\omega,\delta,V,\Omega$ correspond to the natural oscillation frequency, driving frequency, detuning, driving strength and Rabi frequency respectively. We choose a blue detuning $\omega=1.5\omega_0$, and we normalize all the parameters to $\omega_0=1$. We plot the energy spectra and infinite time average energies of the Floquet eigenstates in \cref{Fig: TLS0}. The corresponding analytic formulas are as follows:
	\begin{gather}
		P^\infty_\pm(E)=\frac{\Omega+\delta}{2\Omega}\delta(E-\epsilon_\pm)+\frac{\Omega-\delta}{2\Omega}\delta(E-\epsilon_\pm\mp\omega),\\
		\bar{\epsilon}^\infty_\pm=\mp\frac{1}{2}\qty(\Omega-\frac{\delta\omega}{\Omega})\label{Eq: Two-level average energy}.
	\end{gather}
	
	In this example the eigentriplets $\qty{\epsilon_\pm,\bar{\epsilon}^\infty_\pm,u_\pm(t)}$ are defined and labeled according to the adiabatic continuation with respect to the driving strength $V$. Generally this would not be available due to the ambiguous definition of adiabaticity \cite{Weinberg_2017, Hone_1997}. For the sake of argument we assume that the eigentriplets are still undetermined at the resonance points and we calculate them using \cref{Eq: Eigenspace Initial WF2,Eq: Eigenspace Floquet Function}. At the first resonance $V=V_1$ (such that $\Omega=\omega$), we expand the infinite time average energy matrix in the basis of \cref{Eq: Two-level eigen 0,Eq: Two-level eigen 1}.
	\begin{gather}
		\ket{\Psi(t)}=C_+\mqty(\sqrt{\frac{2\omega-\omega_0}{2\omega}}\\-\sqrt{\frac{\omega_0}{2\omega}}e^{+i\omega t})
		+C_-\mqty(\sqrt{\frac{\omega_0}{2\omega}}e^{-i\omega t}\\\sqrt{\frac{2\omega-\omega_0}{2\omega}})\label{Eq: Two-level eigen ressonance},\\
		\bar{E}^\infty[\Psi]=\mqty(C_+\\C_-)^\dagger\mqty[-\frac{\omega_0}{2}&0\\0&\frac{\omega_0}{2}]\mqty(C_+\\C_-)\label{Eq: Two-level average matrix}.
	\end{gather}
	
	The infinite time average energy matrix is already diagonalized, so the Floquet eigenbasis defined from \cref{Eq: Eigenspace Floquet Function} coincide with the adiabtically connected one (\cref{Eq: Two-level eigen 0,Eq: Two-level eigen 1}). We can better see the effect of this diagonalization in the eigenstate energy spectrum (\cref{Fig: Resonance1}), where we vary the definition of the Floquet eigenfunction by mixing the Floquet eigenfunctions with quasi-energy $\epsilon=0$, or equivalently varying the initial wavefunction $\Psi(0)=u(0)$ in \cref{Eq: Two-level eigen ressonance}. 
	\begin{gather}
		\ket{u_{+\theta}(0)}=\mqty(\cos\theta\\\sin\theta)\qquad
		\ket{u_{-\theta}(0)}=\mqty(\sin\theta\\-\cos\theta)\label{Eq: Initial WF variable definition}.
	\end{gather}
The average energy minimum and maximum occur at the eigenfunctions of the matrix \cref{Eq: Two-level average matrix}, which are equivalent to the adiabatically connected ones. We note that at these points the spectra are maximally separated from each other.
	
\section{Infinitesimal perturbation problem\label{Sec: Avoided crossing}}
	So far we have a method of defining and calculating the Floquet eigenstates of an ideal Hamiltonian using the infinite time average energy, including the resonant region. However, within this framework, the eigenstates, propagators, etc. are sensitive to perturbations, particularly at and near resonance (similar to the perturbation problem of static degenerate systems). As a result, we cannot confidently model real systems because of the inevitable computational or modeling errors, even though the infinitesimally small perturbations should not have any measurable effects. Thus we have to refine the eigenstate definition to reliably model the real systems regardless of any infinitesimal perturbation.
	
	In real experiments, one of the fundamental limitations of any time-dependent quantum system is the observable timescale, which we will use to quantitatively define and assess the accuracy of the Floquet eigenstates. Associated with this observable timescale is a minimum energy resolution $\xi$, which from here on is the desired degree of accuracy and for simplicity we imply its appropriate dimension transformations in upcoming equations.
	
	To better understand the effect of the infinitesimal perturbation, we assume that the real/unperturbed Hamiltonian $H^0(t)$ are known, along with the eigenstates we wish to approximate $\qty{\epsilon^0_n,\bar{\epsilon}^0_n,u^0_n(t)}$.
	\begin{equation}
		\qty\big[\hat{H}^0(t)-i\partial_t]\ket{u_n^0(t)}=\epsilon_n^0\ket{u_n^0(t)}.
	\end{equation}
	
	A model/calculated Hamiltonian $H(t)$ is riddled with infinitesimal undesired perturbations/errors $v$, and in principle, it is hard to distinguish between the perturbation and the actual components of the real Hamiltonian $H^0$. So the only computationally accessible solutions would be these perturbed Floquet eigenstates $\qty{\epsilon_n,\bar{\epsilon}_n,u_n(t)}$, corresponding to the following Floquet Schr\"{o}dinger equation.
	\begin{equation}
		\qty\big[\hat{H}^0(t)+\hat{v}(t)-i\partial_t]\ket{u_n(t)}=\epsilon_n\ket{u_n(t)}\label{Eq: Exact Floquet eq}.
	\end{equation}
	
	For simplicity, we will only consider the weak static perturbation $\abs{v}\ll\xi\ll\abs{H^0},\omega$, although these methods can be generalized for other perturbation methods, e.g. high frequency expansions \cite{Mikami_2016, Casas_2001}.
	
	At the near resonance regime ($\text{mod}(\epsilon_a-\epsilon_b,\omega)<\xi$), we can use the degenerate perturbation theory method to derive the perturbed eigenstates\cite{Eckardt_2015, Hausinger_2010}. In this case the quasi-energy difference is bounded ($\abs{\epsilon-\epsilon^0}=\order{v}$), while the eigenfunctions $u(t)$, their energy spectrum $P^\infty(E)$ and infinite time average energy $\bar{\epsilon}^\infty$ can differ drastically. Even when the perturbation becomes infinitesimal $\abs{v}\ll\xi$, these definition difference can be beyond the acceptable resolution $\xi$. Thus the infinite time average energy method becomes unstable and impractical for realistic applications, e.g. defining a ground state.
	
	On the other hand the perturbation has minimal observable effects if we limit ourselves to small enough time-scales $t<T^U_{max}$. Using the interaction picture we can quantify the differences between the model/perturbed and real/unperturbed systems. The propagators $U(t)$ are sufficient to characterize any observable difference, and the interaction picture propagator $U_I(t)$ extracts this difference by definition.
	\begin{align}
		\hat{U}_I(t,0)=&\mathds{1}-i\int_{0}^{t}\hat{v}_I(\tau)\hat{U}_I(\tau,0)\dd{\tau}\label{Eq: Interaction picture propagator}\\
		\approx&\mathds{1}+\order{\xi}\quad\forall\; t<T^U_{\rm max}.
	\end{align}
	
	We can ignore the effects of non-resonant elements ($\text{mod}(\epsilon_n-\epsilon_m,\omega)>\xi$), as the interaction picture propagator does not diverge from unity at any time-scales.
	\begin{equation}
		\abs{U_{I,mn}(t)-\delta_{mn}}<\order{\xi}\qquad\forall\; t\quad if\nexists\;\abs{\epsilon_m-\epsilon_n}<\xi.
	\end{equation}
	As for the resonant parts we can express it using the transformation matrix $\mathcal{U}(t)$:
	\begin{gather}
		\hat{\mathcal{U}}(t)=\mqty[c_{am}(t)&c_{an}(t)\\c_{bm}(t)&c_{bn}(t)],\\
		\hat{U}_I(t)\approx\hat{\mathcal{U}}(t)\mqty[e^{-ivt}&0\\0&e^{+ivt}]\hat{\mathcal{U}}^\dagger(t).
	\end{gather}
	In this form we can see that $U_I(t)$ only starts to diverge from unity at timescales $t\gtrsim(\order{1/\abs{v}}$, since the transformation matrix is unitary at any give time $\mathcal{UU}^\dagger=\mathds{1}$. We can define this as the timescale $T^U_{max}$, up to which any observable of the model and real system are indistinguishable.
	
	Thus the model can approximate the real system for weak enough perturbations $\abs{v}\ll\xi$, and we want to find an eigenstate definition which can describe both systems within the timescale limitations $t<T^U_{max}$. Since the infinite time method goes beyond this limit, we have to choose a different quantum number to describe the observable effects.
	
	\subsection{Observed average energy\label{Sec: Approximate average energy}}
	We define the observed average energy as the average energy expectation value up to a finite time $\mathcal{T}$ (\cref{Eq: Average energy def}), within reasonable experimental timescales $\mathcal{T}<T^U_{max}$. Our goal is to find a definition of the average energy functional $\bar{E}^\mathcal{T}$ which closely approximates the real/unperturbed one $\bar{E}^{0\mathcal{T}}$ within a predefined acceptable accuracy range $\xi$. With this we can trivially approximate the unperturbed eigenstates using a variational principle with tolerance $\xi$.
	\begin{gather}
		\bar{E}^\mathcal{T}[\Psi(0)]=\eev{\hat{U}^\dagger\hat{H}\hat{U}}{\Psi(0)}_\mathcal{T}\label{Eq: Average energy condition},\\
		\abs{\bar{E}^\mathcal{T}[\Psi]-\bar{E}^{0\mathcal{T}}[\Psi]}<\xi\qquad\forall\;\Psi(0)\label{Eq: Average energy condition2}.
	\end{gather}
	
	For simplicity we assume the real system does not have near resonance pairs, only exact resonances and far from resonance states. In this case the real system is well described by the unperturbed infinite time eigenstates $\mathcal{E}^\infty_{H^0}$, and the unperturbed observed average energy quickly converges to the infinite time one.
	\begin{gather}
		\mod(\epsilon^0_n-\epsilon^0_m,\omega)\in\qty{0,\gg\xi}\qquad\forall\;m,n,\\
		\abs{\bar{E}^{0\mathcal{T}}[\Psi]-\bar{E}^{0\infty}[\Psi]}<\xi\qquad\forall \mathcal{T}>T^{\bar{E}}_{min}.
	\end{gather}
	
	Including near resonances in the real/unperturbed system, the observed average energy and its eigenstates can still be defined for the real system by reapplying the discussions in this section, in which case the observed unperturbed eigenstates $\mathcal{E}^{\mathcal{T}}_{H^0}$ would be the target we wish to derive instead of the infinite time solutions $\mathcal{E}^\infty_{H^0}$. We explain more in \cref{Sec: Real near resonance}, and here we only concentrate on deriving the infinite time solutions of the real/unperturbed system through the observed average energy.
	
	Decomposing the definition in \cref{Eq: Average energy condition} using the unperturbed eigenstates $\qty{\epsilon^0_n,\bar{\epsilon}^{0\infty}_n,u^0_n(t)}$, we find two timescale boundaries $T^{\bar{E}}_{min}$ and $T^{\bar{E}}_{max}$ within which the observed average energy satisfies:
	\begin{equation}
		\bar{E}^{\mathcal{T}}=\sum_{n}\abs{C^0_n}^2\bar{\epsilon}^{0\mathcal{T}}_n+\order{\xi}\qquad\forall\; T^{\bar{E}}_{min}<\mathcal{T}<T^{\bar{E}}_{max}.
	\end{equation}
	The derivation of these boundaries and related discussions are presented in \cref{Sec: Average energy boundary}. Roughly these boundaries are related to the timescale where the unperturbed eigenstates become relevant $\mathcal{T}>T^{\bar{E}}_{min}$, and the timescale before which the infinitesimal perturbations $v$ can be ignored $\mathcal{T}<T^U_{max}\lesssim T^{\bar{E}}_{max}$. These boundaries only become relevant in further theoretical derivations based on the Floquet average energy, and in practice we are only concerned if a reasonable timescale can be defined within these boundaries. Since for any averaging time within this region, we get an equally good approximation to the exact average energy functional, we can choose an arbitrary timescale to define the appropriate observed average energy functional, and in most cases $\mathcal{T}\approx1/\xi$ would be a good choice for this.
	
	With this definition of observed average energy alone we can derive a close approximation of the unperturbed eigenbasis $\mathcal{E}^0_{H^0}$ from its variational principle around the initial wavefunction $\bar{E}^\mathcal{T}[\Psi(0)]$.
	\begin{gather}
		\bar{\epsilon}^{\mathcal{T}}_n\equiv\qty{\bar{E}^\mathcal{T}[\Psi(0)]\Big\vert \delta\bar{E}^\mathcal{T}[\Psi]<\xi}\label{Eq: Variational method practical3},\\
		\mathcal{E}^{0\xi}_{H}\equiv\qty{\bar{\epsilon}^{\mathcal{T}}_n,\Psi_n(0)\vert\delta\bar{E}^\mathcal{T}[\Psi]<\xi},\\
		\bar{\epsilon}^{\mathcal{T}}_n=\bar{\epsilon}^{0\infty}_n+\order{\xi}\quad\ket{\Psi_n(0)}=\ket{\Psi^0_n(0)}+\order{\xi}\label{Eq: Variational method practical4}.
	\end{gather}
	
	As for the computationally accessible forms, we rely on the observation that the effective average energy $\bar{\epsilon}^T[u]$ in the near resonant susbspace $\mathds{HT}_{\epsilon\xi}$ retain the same properties of \cref{Eq: Effective Average Propert0,Eq: Effective Average Propert1}, up to the acceptable accuracy $\xi$.
	\begin{gather}
		\abs{\epsilon[u]-\epsilon}<\xi\quad\eval{\delta\epsilon[u]}_{\mathds{HT}}<\xi\quad\forall u(t)\in\mathds{HT}_{\epsilon\xi},\\
		\abs{\bar{\epsilon}^T[u_a]-\bar{E}^\mathcal{T}[\Psi_a]}<\xi\quad\forall u_a(t)\in\mathds{HT}_{\epsilon\xi},\\
		\abs\Big{\eval{\delta\bar{\epsilon}^T[u_a]}_{\mathds{HT}_{\epsilon\xi}}-\eval{\delta\bar{E}^\mathcal{T}[\Psi_a]}_{\mathds{HT}_{\epsilon\xi}}}<\xi\quad\forall u_a(t)\in\mathds{HT}_{\epsilon\xi}.
	\end{gather}
	
	The variational procedure is thus analogous to the one presented in \cref{Sec: Exact average energy}, but in a more extended search space.
	\begin{gather}
		\epsilon_n\equiv\qty\Big{\epsilon[u]\Big\vert\eval{\delta\epsilon[u]}_{\mathds{HT}}<\xi}\Rightarrow\mathds{HT}_{\epsilon\xi}\label{Eq: Variational method practical0}\beginsubeqn,\\
		\bar{\epsilon}^{\mathcal{T}}_n\equiv\qty\Big{\bar{\epsilon}^T[u_a]\Big\vert\eval{\delta\bar{\epsilon}^T[u_a]}_{\mathds{HT}_{\epsilon\xi}}<\xi}\subeqn\label{Eq: Variational method practical1},\\
		\mathcal{E}^{\mathcal{T}}_{H}\equiv\qty\Big{\epsilon_n,\bar{\epsilon}^{\mathcal{T}}_n,u_n(t)\Big\vert\delta\epsilon[u]<\xi\rightarrow\eval{\delta\bar{\epsilon}^T[u]}_{\mathds{HT}_{\epsilon_n\xi}}<\xi}\label{Eq: Variational method practical2}.
	\end{gather}
	where \cref{Eq: Variational method practical0} implies the quasi-energy resonance is not lifted if the difference is within the acceptable error $\xi$. This accounts for the infinitesimal resonance lifting previously presented, so that we get consistent solutions near the real/unperturbed eigenstates $\qty{\epsilon^0_n,\bar{\epsilon}^{0\infty}_n,u^0_n(t)}$ for arbitrary weak interaction $\abs{v}\ll\xi$.
	
	The equivalent Lagrange minimization method remains roughly the same, but with appropriate change of Lagrangian multiplier $\lambda_\xi$ to account for the finite resolution of $\delta\epsilon$:
	\begin{equation}
		\bar{\epsilon}^\mathcal{T}_n=\crit_{u(t)}\qty{\bar{\epsilon}_T[u]+\lambda_\xi\fdv{\epsilon[u]}{u}+\bar{\epsilon}(\iip{u}-1)}+\order{\xi}\label{Eq: Variational method practical5}.
	\end{equation}
	
	The same variational procedure are used when the real system has infinitesimal near resonance (\cref{Sec: Real near resonance}). We also point out that, although in these derivations we have assumed we know the exact/unperturbed average energy and eigenstates we wish to approximate, in practice these are not necessary. We can discuss whether the observed average energies and their eigenstates of the model/perturbed system closely apporximate the real system by varying the timescale $\mathcal{T}$ or equivalently the acceptable accuracy $\xi$, and observe the stability of these solutions.
	
	Thus, we have a more robust description of the Floquet eigenbasis $\mathcal{E}^{\mathcal{T}}_{H}=\qty{\epsilon_n,\bar{\epsilon}^{\mathcal{T}}_n,u_n(t)}$, which has the same benefits of the definition in \ref{Sec: Exact average energy}, but also is robust against infinitesimal perturbation. In this form unaccounted perturbations or numerical errors will not break the definition of the basis set.
	
	\subsection{Floquet-Ritz variation principle\label{Sec: Floquet-Ritz}}
	The power of the Ritz variational principle is in the approximation of the ground state on a much smaller Hilbert subspace. For this we have to investigate how the ground state approximation changes as we increase the Hilbert space.
	
	 First it should be noted that in the full extended Hilbert space $\mathds{HT}$, both the infinite time and observed average energies are lower bounded by those of the ground states, $\bar{\epsilon}^{\infty}_0$ or $\bar{\epsilon}^{\mathcal{T}}_0$, respectively. Thus at the limit of the Hilbert space expansion, we have a Ritz-like variation principle.
	\begin{align}
		\bar{E}^\infty[u]=\sum_{n}\abs{C^{\infty}_n}^2\bar{\epsilon}^{\infty}_n\geq\bar{\epsilon}^{\infty}_0\quad&\forall u(t),\\
		\bar{E}^\mathcal{T}[u]=\sum_{n}\abs{C^{\mathcal{T}}_n}^2\bar{\epsilon}^{\mathcal{T}}_n+\order{\xi}\gtrsim\bar{\epsilon}^{\mathcal{T}}_0\quad&\forall u(t).
	\end{align}
	For the latter variation to be applicable, we assume that the two lowest observed average energies are sufficiently well separated $\bar{\epsilon}^{\mathcal{T}}_1-\bar{\epsilon}^{\mathcal{T}}_0\gg\xi$, so that the global minimum gives a good approximation of the ground state $C^{\mathcal{T}}_0=1+\order{\xi/(\bar{\epsilon}^{\mathcal{T}}_1-\bar{\epsilon}^{\mathcal{T}}_0)}$.
	
	In order to have a proper Ritz variational principle, we assume we have a well behaved expansion series of the Hilbert space $\mathds{HT}_i$, e.g. one derived from a Davidson algorithm\cite{Davidson_1975, Sahoo_2019}, and we can find a lower-bound in the expansion $i$ that closely approximates the exact solution. In other words, for a given accuracy $\xi$, the ground state $\eval{\qty{\epsilon_0,\bar{\epsilon}_0,u_0}}_{\mathds{HT}_i}$, or equivalently the average energy functional $\eval{\bar{E}[u]}_{\mathds{HT}_i}$, evaluated in the Hilbert subspace $\mathds{HT}_i$ are within the acceptable accuracy $\xi$ of their exact counterparts, and subsequent expansions will not improve the accuracy more than that.
	\begin{gather}
		\eval{\qty{\epsilon_0,\bar{\epsilon}_0,u_0}}_{\mathds{HT}_i}=\eval{\qty{\epsilon_0,\bar{\epsilon}_0,u_0}}_{\mathds{HT}}+\order{\xi},\\
		\eval{\bar{E}[u]}_{\mathds{HT}_i}=\eval{\bar{E}[u]}_{\mathds{HT}}+\order{\xi}.
	\end{gather}
	
	In the simplest case we can model such a Hilbert space expansion with a monotonic decrease in the coupling $v^i$ between two Hilbert subspaces, $\mathds{HT}_i$ and its complement $\mathds{HT}_{i\perp}$, corresponding to the Hamiltonians $H^i$ and $H^{i\perp}$ respectively. The Hilbert subspace $\mathds{HT}_i$ is where the current ground state approximation resides, and increases after each iteration. The full Floquet Hamiltonian can thus be decomposed as follows at any given step $i$ in the expansion.
	\begin{gather}
		\hat{H}^F(t)=\mqty[\hat{H}^{i\perp}(t)&\hat{v}^i(t)\\\hat{v}^{i\dagger}(t)&\hat{H}^i(t)]+\mqty[-i\partial_t&0\\0&-i\partial_t]\label{Eq: Floquet Hamiltonian subspace},\\
		\abs{\hat{v}^{i+1}}\leq\abs{\hat{v}^{i}}\qquad \mathds{HT}_{i+1}\supset\mathds{HT}_{i}\label{Eq: Floquet Ritz expansion condition}.
	\end{gather}
	
	Whether a given algorithm does yield such a property is still up for debate. For now we are only concerned if such an algorithm can yield a Ritz-like variational principle and what ground state definition it should follow.
	
	First let's consider the behavior at a weak coupling threshold $\abs{v^{i_c}}\lesssim\xi$. In this case we can use the previous arguments in \cref{Sec: Approximate average energy} to find that the observed average energy ground state of the full Hilbert space $\eval{\qty(\epsilon^{\mathcal{T}}_0,\bar{\epsilon}^{\mathcal{T}}_0,u^{\mathcal{T}}_0)}_{\mathds{HT}}\in\mathcal{E}^\xi_H$ is closely approximated by the decoupled solution of \cref{Eq: Floquet Hamiltonian subspace}:
	\begin{equation}
		\mqty[\hat{H}^{i_c\perp}(t)&0\\0&\hat{H}^{i_c}(t)]+\mqty[-i\partial_t&0\\0&-i\partial_t],
	\end{equation}
	which in this case is the observed ground state $\eval{\qty(\epsilon^{\mathcal{T}i_c}_0,\bar{\epsilon}^{\mathcal{T}i_c}_0,u^{\mathcal{T}i_c}_0)}_{\mathds{HT}_{i_c}}$ of the Hamiltonian $H^{i_c}(t)$ in the limited Hilbert space $\mathds{HT}_{i_c}$, assuming the correct subspace is chosen. Similarly all of the expansion points above this threshold $i>i_c$ have the same property, as long as the expansion series satisfies \cref{Eq: Floquet Ritz expansion condition}. We can simplify this statement using the observed average energy functional evaluated in the truncated Hilbert space $\mathds{HT}_i$, compared to the exact one evaluated in the full Hilbert space $\mathds{HT}$:
	\begin{equation}
		\eval{\bar{E}^{\mathcal{T}}[u]}_{\mathds{HT}_i}=\eval{\bar{E}^{\mathcal{T}}[u]}_{\mathds{HT}}+\order{\xi}\quad\forall \mathds{HT}_i\supset\mathds{HT}_{i_c}.
	\end{equation}
	
	The same is not true for the infinite time ground state definition, where we can find cases where the asymptotic limit does not correspond to the full Hilbert space solution, provided that such a limit can even be found.
	\begin{equation}
		\eval{\bar{E}^{\infty}[u]}_{\mathds{HT}_{i_c}}\neq\lim_{i\to\infty}\eval{\bar{E}^{\infty}[u]}_{\mathds{HT}_i}\neq\eval{\bar{E}^{\infty}[u]}_{\mathds{HT}}.
	\end{equation}
	
	Therefore a Floquet Ritz variational principle can be formulated based on the observed ground state definition. The condition for the convergence and the Ritz inequality are as follows:
	\begin{gather}
		\eval{\delta\epsilon[u^{\mathcal{T}i}_0]}_{\mathds{HT}}<\xi\;\text{and}\;\eval{\delta\bar{E}^\mathcal{T}[u^{\mathcal{T}i}_0]}_{\mathds{HT}}<\xi\label{Eq: Condition Floquet Ritz},\\
		\bar{\epsilon}^{\mathcal{T}i}_0\in\qty{\bar{\epsilon}^{\mathcal{T}}_n+\order{\xi}}\geq\bar{\epsilon}^{\mathcal{T}}_0+\order{\xi}\label{Eq: Ritz variation}.
	\end{gather}
	
	The Floquet Ritz method has similar properties to the static method. If the initial guess is not good enough we risk convergence to an excited state rather than the desired ground state (\cref{Eq: Ritz variation}).
	
	There is however a caveat to this method. The functional used in evaluating the average energy $\eval{\bar{E}^{\mathcal{T}}}_{\mathds{HT}_{i'}}$ at intermediate steps, namely the effective average energy functional $\eval{\bar{\epsilon}^T[u]}_{\mathds{HT}_{i'}}$, or the propagator $\eval{U(t)}_{\mathds{HT}_{i'}}$, do not necessarily give a good approximation to the average energy functional in the full Hilbert space $\eval{\bar{E}^\mathcal{T}}_{\mathds{HT}}$. It is thus possible to find "average energies" well below the observed ground state average energy in the full Hilbert space.
	\begin{equation}
		\exists\;\qty\Big{i'<i_c\Big\vert\eval{\bar{\epsilon}^{\mathcal{T}i'}_0}_{\mathds{HT}_{i'}}<\eval{\bar{\epsilon}^{\mathcal{T}}_0}_{\mathds{HT}}+\order{\xi}}.
	\end{equation}
	
	This does not contradict the Ritz variational principle in \cref{Eq: Ritz variation}, since the solution at these points do not satisfy \cref{Eq: Condition Floquet Ritz} and are not convergent yet. However this can still create problems depending on the algorithm used to approach the convergence. E.g. if we select the ground state at each step to determine subsequent Hilbert spaces $\mathds{HT}_{i+1}$, it is possible to flip-flop between approximations of various different eigenstates $(\epsilon^{\mathcal{T}}_n,\bar{\epsilon}^{\mathcal{T}}_n,u^{\mathcal{T}}_n)$, and even to converge to an excited state despite starting from a good initial guess.
	
	Nevertheless, in principle a Floquet-Ritz variational principle is possible on limited Hilbert spaces $\mathds{HT}_i$, as long as we use the observed average energy labeling, either explicitly or implicitly. This is the main point we want to state regarding a possible Floquet-Ritz theory.
	
	\subsection{Perturbed two-level system\label{Sec: Example perturbed two-level}}
	\begin{figure*}
		\hfill\subfloat[\label{Fig: Pert TLS}]{%
			\subfigimg[scale=1.]{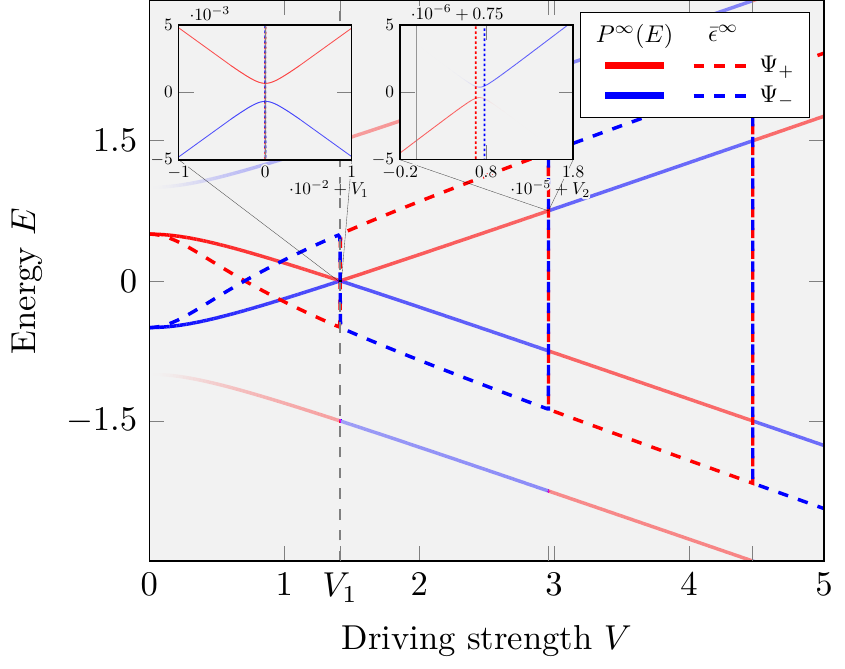}{10pt}{1.5}%
		}\hfill
		\subfloat[\label{Fig: Pert Resonance1}]{%
			\subfigimg[scale=1.]{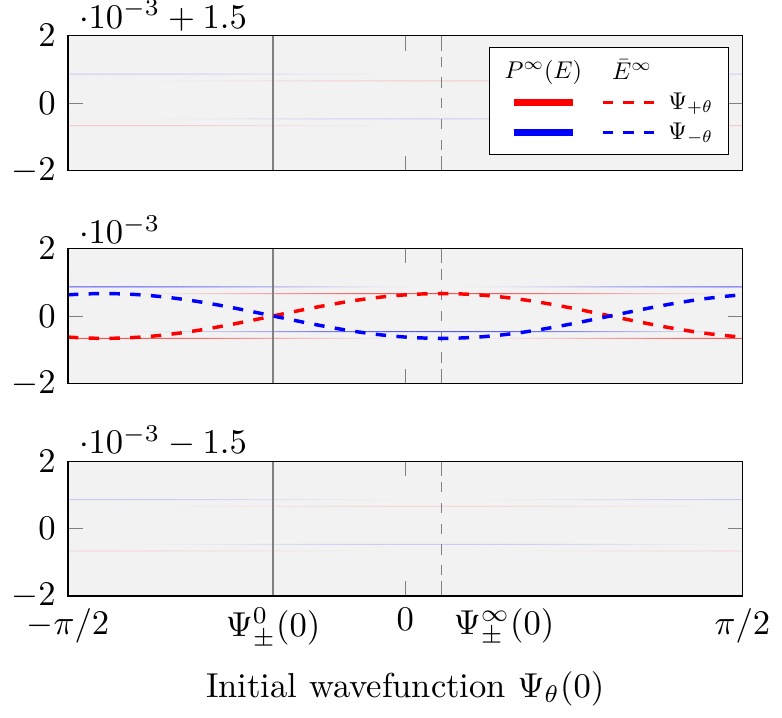}{25pt}{2.25}%
		}\hfill
		\caption{Equivalent energy spectrum and infinite time average energy plot to \cref{Fig: TLS0} for the perturbed two-level system. The eigenstates are determined exactly from \cref{Eq: Pert Two-level exact eigen} and labeled adiabatically. The vertical dashed lines correspond to the infinite time eigenstate solutions at the first avoided crossing $V_1$, where both plots correspond to each other. The vertical solid line are the eigenstate solutions of the unperturbed system (\cref{Fig: Resonance1}).\label{Fig: Pert TLS0}}
	\end{figure*}
	A minimal example including the perturbation effect presented above is the two-level system in \cref{Eq: Two-level Hamiltonian} perturbed by a weak static coupling. Physically this can be regarded as a stray static electric field contamination or computationally it can be a numerical error.
	\begin{equation}
		\hat{H}(t)=\mqty[\frac{\omega_0}{2}&v+\frac{V}{2}e^{-i\omega t}\\v+\frac{V}{2}e^{+i\omega t}&-\frac{\omega_0}{2}].
	\end{equation}
	
	 For consistency we consider the unperturbed Hamiltonian (\cref{Eq: Two-level Hamiltonian}) to be the real system and we project the perturbed Floquet Hamiltonian on the unperturbed Floquet eigenfunctions (\cref{Eq: Two-level eigen 0,Eq: Two-level eigen 1}):
	\begin{gather}
		\hat{H}^F(t)=\mqty[\epsilon^0_+&0\\0&\epsilon^0_-]+v\mqty[-\frac{V}{\Omega}\cos\omega t&\frac{\Omega+\delta}{2\Omega}-\frac{\Omega-\delta}{2\Omega}e^{-i2\omega t}\\
		\frac{\Omega+\delta}{2\Omega}-\frac{\Omega-\delta}{2\Omega}e^{+i2\omega t}&\frac{V}{\Omega}\cos\omega t]-i\partial_t\label{Eq: Projected Floq Hamil},\\
		\ket{\Psi_\pm(t)}=e^{-i\epsilon_\pm t}\hat{\mathcal{U}}^0(t)\mqty(u_{+\pm}(t)\\u_{-\pm}(t)),\\
		\mathcal{U}^0(t)=\mqty[\sqrt{\frac{\Omega+\delta}{2\Omega}}&\sqrt{\frac{\Omega-\delta}{2\Omega}}e^{-i\omega t}\\
			-\sqrt{\frac{\Omega-\delta}{2\Omega}}e^{+i\omega t}&\sqrt{\frac{\Omega+\delta}{2\Omega}}].
	\end{gather}
	
	We will focus on the first resonance point $\Omega=\omega$. Here the driving frequency $\omega$ is much lager than the static components ($\epsilon^0_+-\epsilon^0_-=0\ll\omega$), so we can approximate the Hamiltonian by ignoring the time-dependent components in the projected Floquet Hamiltonian (\cref{Eq: Projected Floq Hamil}). The exact perturbed Floquet eigenstates can then be approximated to:
	\begin{gather}
		\hat{H}^F(t)\approx v\frac{2\omega-\omega_0}{2\omega}\mqty[0&1\\1&0]-i\partial_t,\\
		\ket{\Psi_\pm(t)}\approx e^{-i\epsilon_\pm t}\mqty[\sqrt{\frac{2\omega-\omega_0}{2\omega}}&\sqrt{\frac{\omega_0}{2\omega}}e^{-i\omega t}\\
			-\sqrt{\frac{\omega_0}{2\omega}}e^{+i\omega t}&\sqrt{\frac{2\omega-\omega_0}{2\omega}}]\mqty(\frac{1}{\sqrt{2}}\\\pm\frac{1}{\sqrt{2}})\label{Eq: Pert Two-level exact eigen},\\
		\epsilon_\pm\approx\pm v\frac{2\omega-\omega_0}{2\omega}.
	\end{gather}
	These eigenstates correspond to the infinite time eigenstates with the average energies almost coinciding with each other.
	\begin{gather}
		\bar{\epsilon}^\infty_\pm\approx\pm v\frac{2\omega-\omega_0}{2\omega},\\
		\abs{\bar{\epsilon}^\infty_\pm}\sim\frac{v}{2}\ll\abs{\bar{\epsilon}^0_\pm}=\frac{\omega_0}{2}.
	\end{gather}
	
		Here we see the limitations of the infinite time average energy method. As we approach the limit $v\to 0$, the infinite time average energies difference of the model/perturbed systems vanishes $\bar{\epsilon}^\infty_+-\bar{\epsilon}^\infty_-\to 0$, while the real/unperturbed solution at $v=0$ does not ($\bar{\epsilon}^0_+-\bar{\epsilon}^0_-= \omega_0\neq0$). Such mismatch in energies does not occur in the static case and is specific to the resonance of Floquet systems. We also see that the eigenstates (\cref{Eq: Pert Two-level exact eigen}) do not change as we approach the limit $v\to0$, even as they are defined by the infinite time average energy.
	
	We plot the infinite time Floquet eigenstates and average energies in \cref{Fig: Pert TLS0} with the perturbation exaggerated to $v=10^{-3}\omega_0$ for clarity. This system is simple enough that we can compute the exact propagators.
	\begin{align}
		\mathcal{U}^0(t)=&\mqty[\sqrt{\frac{2\omega-\omega_0}{2\omega}}&\sqrt{\frac{\omega_0}{2\omega}}e^{-i\omega t}\\
			-\sqrt{\frac{\omega_0}{2\omega}}e^{+i\omega t}&\sqrt{\frac{2\omega-\omega_0}{2\omega}}],\\
		U(t)=&\mathcal{U}^0(t)
			\mqty[\cos\epsilon_+t&-i\sin\epsilon_+t\\
			i\sin\epsilon_+t&\cos\epsilon_+t]
			\mathcal{U}^{0\dagger}(0)\label{Eq: Pert prop two-level},\\
		U^0(t)=&\mathcal{U}^0(t)\mathcal{U}^{0\dagger}(0)\label{Eq: unPert prop two-level}.
	\end{align}
	
	As shown in \cref{Fig: Pert TLS}, the infinite time eigenstate solutions of the perturbed system are closely approximating the unperturbed eigensolutions (\cref{Fig: TLS}) at all the non-resonant points, with the exception of a trivial label swap caused by the difference in the adiabatic continuations. However at the resonance point these solutions differ drastically, which is best seen in the resonance energy spectrum (\cref{Fig: Pert Resonance1}) plotted against the initial wavefunction with the definition given in the previous example (\cref{Eq: Initial WF variable definition}). Here we notice that the range of the average energy variation is much smaller than that of the unperturbed case (fig.2b). Also, the eigenstates of the unperturbed and perturbed systems correspond to different values of $\theta$ (dashed and continuous gray line). Thus we confirm that the labeling of the Floquet eigenstates by the infinite time average energy can be inconsistent with the inclusion of infinitesimal perturbations.
	
	\begin{figure*}
		\hfill\subfloat[\label{Fig: Energy landscape}]{%
			\subfigimg[scale=1.1]{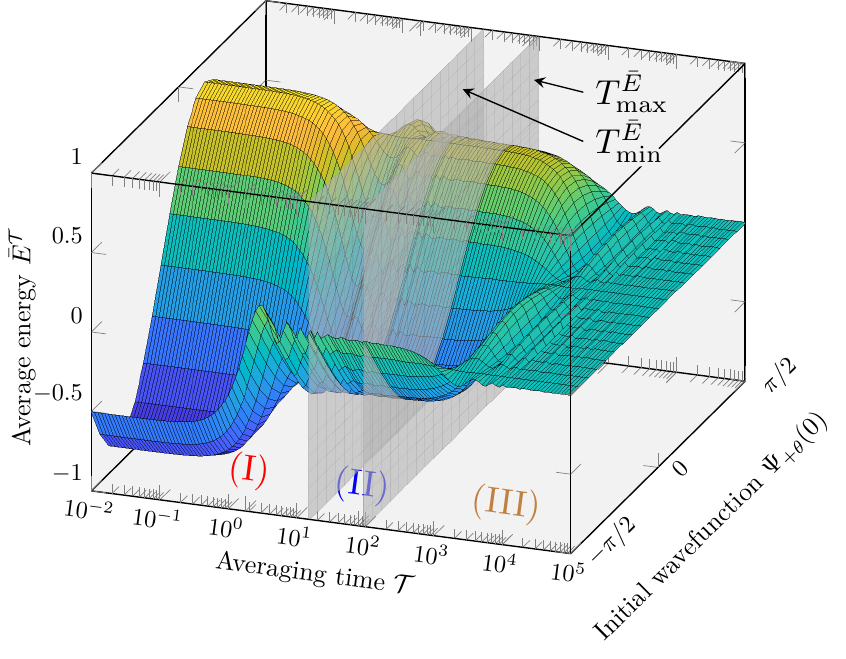}{10pt}{1.5}
		}\hfill
		\subfloat[\label{Fig: Energy variation}]{%
			\subfigimg[scale=.9]{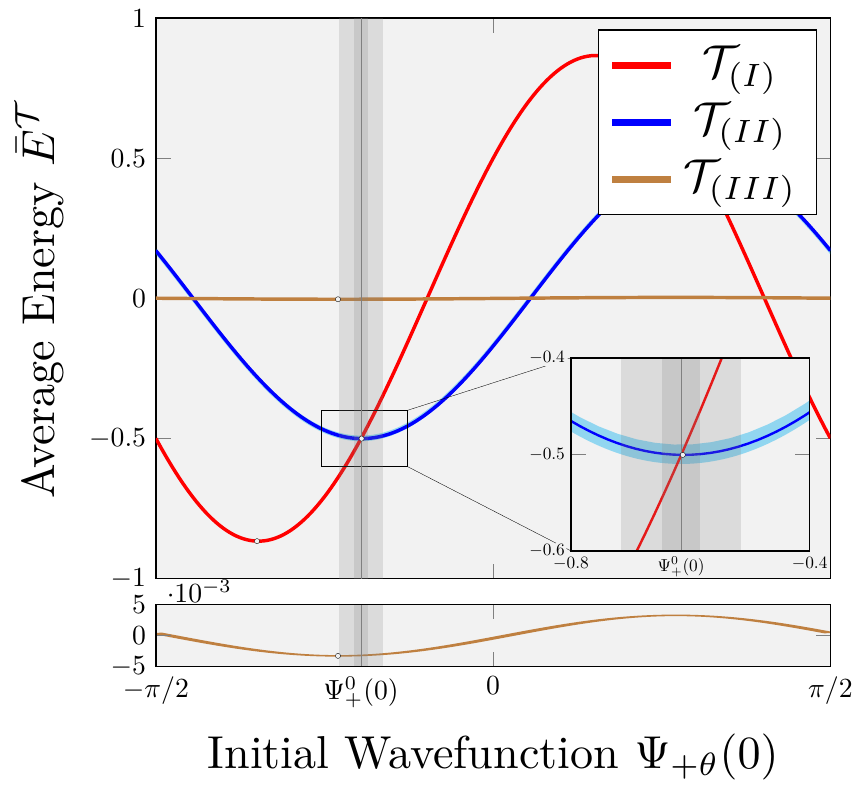}{10pt}{1.5}
		}\hfill
		\caption{\protect\subref{Fig: Energy landscape} Observed average energy landscape $\bar{E}^{\mathcal{T}}$ of the perturbed two-level system. The three distinct regions correspond to (I) the instantaneous energy, (II) unperturbed average energy, (III) perturbed infinite time average energy. The planes $T^{\bar{E}}_{min}$ and $T^{\bar{E}}_{max}$ delimit the region where the approximate average energy approximate the unperturbed one within the acceptable accuracy $\bar{E}^{\mathcal{T}}=\bar{E}^{0}+\order{\xi}$. The acceptable accuracy is exaggerated to $10^{-2}\omega_0$. The plots in \protect\subref{Fig: Energy variation} are taken with averaging times $\mathcal{T}$ well within the defined regions ($\mathcal{T}_{(I)}=10^{-1.5},\,\mathcal{T}_{(II)}=10^{1.5},\,\mathcal{T}_{(III)}=10^{4.5}$). The cyan region in the zoomed in plot of \protect\subref{Fig: Energy variation} represent the acceptable error of the unperturbed average energy $\bar{E}^{0\infty}\pm\xi$. The vertical gray regions correspond to the acceptable deviation from the unperturbed eigenstate with acceptable accuracies $\abs{v}$ and $\xi$ corresponding to the inner and outer region respectively.\label{Fig: Energy landscapes}}
	\end{figure*}
	Having shown the limitations of the infinite time approach, now we compare it to the observed average energy method. First we want to confirm that our original premise for deriving the observed eigenstates is valid, i.e. any observable of the perturbed and unperturbed system can be within an acceptable error at some timescale. Since the exact propagators are known at the first resonance point (\cref{Eq: Pert prop two-level,Eq: unPert prop two-level}), we can directly find the timescales $\mathcal{T}$ where propagators in the real and model systems are equivalent, for arbitrarily chosen accuracy $\xi$. For our purposes it is sufficient to expand the model propagator $U(t)$ up to first order in $v$ and derive the relevant timescale $\mathcal{T}$ from there.
	\begin{gather}
		\abs{U(t)-U^0(t)}\sim\frac{(2\omega-\omega_0)v}{2\omega}t+\order{v^2}<\xi/\omega_0,\\
		\mathcal{T}<T^U_{max}\sim\frac{2\omega\xi}{(2\omega-\omega_0)\omega_0 v}.
	\end{gather}
	
	As long as the observation timescale and the timescale for determining the average energy are within this range, we can conclude that the model/perturbed system does indeed well approximate the real/unperturbed system. Here we have used the dimension transformation of $\xi$ with the typical energy scale as $\omega_0$. For the two-level system this rough approximation is sufficient, especially since the most interesting physical system is where all of the interactions are of the same order $\order{\omega}=\order{\omega_0}=\order{V}$ and the time-periodic interactions are most prominent.
	
	Within these timescales $\mathcal{T}$ and acceptable accuracies $\xi$, we can derive the observed average energy functional $\bar{E}^{\mathcal{T}}$ and its resulting eigenstates directly from \cref{Eq: Average energy def,Eq: Variational method practical3}. But first we look at the observed average energies boundaries ($T^{\bar{E}}_{min}$ and $T^{\bar{E}}_{max}$), which we have proposed to determine the region where the model eigenstates approximate the real ones. Since we know the exact Hamiltonians, we can directly derive them from \cref{Eq: Def min boundary,Eq: Def max boundary}:
	\begin{align}
		T^{\bar{E}}_{min}=&\frac{V}{\omega\xi}&T^{\bar{E}}_{max}=&\frac{2\omega\xi}{V^2v}\approx T^U_{max}.
	\end{align}
	We see here a good agreement between the two higher bounds ($T^{\bar{E}}_{max}$ and $T^U_{max}$) considering that $\order{\omega}=\order{\omega_0}=\order{V}$, in accord with our derivation. The lower boundary in this system, assures that the observed average energy functional approximates the infinite time average energy of the unperturbed system, and not another functional with different critical points.
	
	In the limit of $v\to0$ we find the higher bounds diverge to infinity, suggesting that at arbitrary timescales, we would not be able to find an observable difference between the model and real system, perfectly in accord with the physical intuition. Comparing the lower and higher bounds ($T^{\bar{E}}_{min}$ and $T^{\bar{E}}_{max}$), we find that for small perturbations $v$ we can always find an acceptable error $\xi$, up to which the model Hamiltonian approximates the real one, although this is often too large for practical applications.
	\begin{equation}
		T^{\bar{E}}_{max}>T^{\bar{E}}_{min}\Leftrightarrow\xi>\order{\omega\sqrt{v/\omega}}>v.
	\end{equation}
	
	For this example system we choose the acceptable accuracy to be $\xi=10^{-2}\omega_0$ so that we can define an observable timescale between the boundaries. The effect of these boundaries are best seen in \cref{Fig: Energy landscapes}, where the energy functional $\bar{E}^{\mathcal{T}}[\Psi(0)]$ is evaluated for varying timescales resulting into three distinct regions. These regions represent the timescale where the energy functional $\bar{E}^{\mathcal{T}}[\Psi(0)]$ approximates: (I) the instantaneous energy; (II) observed/unperturbed infinite time average energy; (III) perturbed infinite time average energy. We can see the accuracy of these descriptions in \cref{Fig: Energy variation}, as well as the deviation from the acceptable regime outside the boundaries (with the exception of region (III), which has not yet converged due to numerical limitations).
	\begin{align}
		\bar{E}^{\mathcal{T}}_{(I)}[\Psi]&\approx \ev{\hat{H}(0)}{\Psi(0)},\\
		\bar{E}^{\mathcal{T}}_{(II)}[\Psi]&\approx \bar{E}^0[\Psi]+\order{\xi},\\
		\bar{E}^{\mathcal{T}}_{(III)}[\Psi]&\approx \bar{E}^{\infty}[\Psi].
	\end{align}
	
	\begin{figure}
		\includegraphics[width=8.6cm]{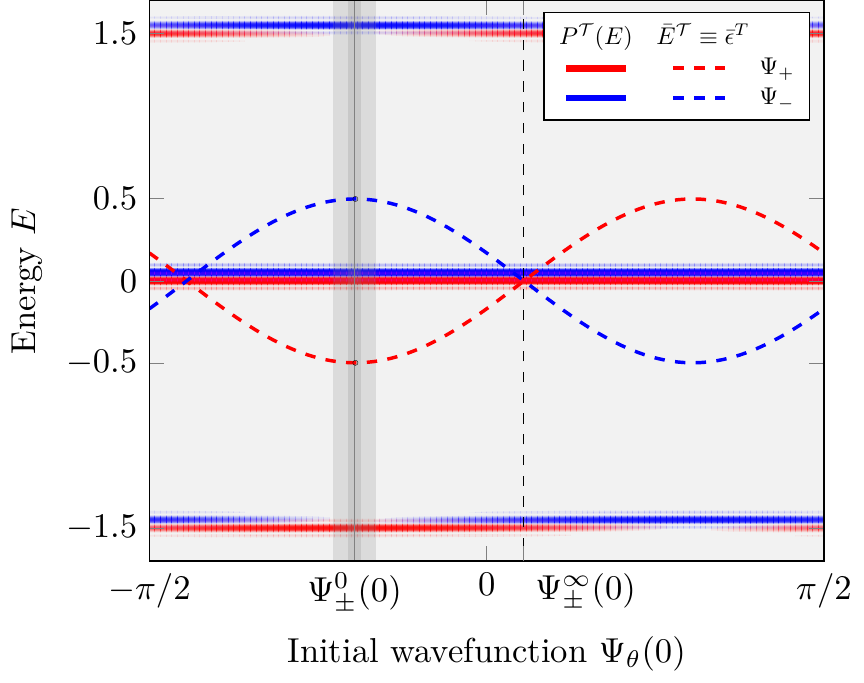}
		\caption{Observed energy spectrum $P^{\mathcal{T}}(E)$ and average energy $\bar{E}^{\mathcal{T}}$/effective average energy $\bar{\epsilon}^T$ of the perturbed two-level system with $\mathcal{T}=10^{-2}\omega_0$. The vertical gray regions are equivalent to the ones in \cref{Fig: Energy variation}, and the dashed vertical line correspond to the infinite time eigenstate solutions.\label{Fig: Pert Resonance2}}
	\end{figure}
	
	Next we look at the derivation of the observed average energies and the eigenstates. Using the unperturbed Floquet basis, we can compute the exact quasi-energy, observed average energy and effective average energy at the near resonance for different Floquet functions $u_\theta(t)$, other than the eigenstates. For sake of simplifying the equation, the Floquet functions are constructed by mixing the unperturbed Floquet eigenfunctions, which we wish to derive.
	\begin{gather}
		\ket{u_\theta(t)}=\cos\theta\ket{u_+^0(t)}+\sin\theta\ket{u_-^0(t)},\\
		\epsilon(\theta)=\frac{(2\omega-\omega_0)v}{2\omega}\sin2\theta<\xi\quad\forall\theta\label{Eq: Two-level quasi variation}.
	\end{gather}
	\begin{align}
		\bar{\epsilon}^{T}(\theta)=&\mqty(\cos\theta\\\sin\theta)^\dagger\mqty[-\frac{\omega_0}{2}&\frac{(2\omega-\omega_0)v}{2\omega}\\\frac{(2\omega-\omega_0)v}{2\omega}&\frac{\omega_0}{2}]\mqty(\cos\theta\\\sin\theta)\label{Eq: Two-level effective average}\\
		\approx&\bar{E}^{\mathcal{T}}[u_\theta]=\mqty(\cos\theta\\\sin\theta)^\dagger\mqty[-\frac{\omega_0}{2}&0\\0&\frac{\omega_0}{2}]\mqty(\cos\theta\\\sin\theta)+\order{\xi}.
	\end{align}
	
	We see that the effective average energy $\bar{\epsilon}^T$ is indeed a close approximation to unperturbed average energy $\bar{E}^0$ up to the accuracy of $\xi$. Since the unperturbed average energies are well separated $\bar{\epsilon}^0_+-\bar{\epsilon}^0_0\gg\xi$, and the quasi-energies are relatively stationary (\cref{Eq: Two-level quasi variation}), we can get a close approximation to the unperturbed basis set from the variation of \cref{Eq: Two-level effective average}. We plot this observed average energy in \cref{Fig: Pert Resonance2}, where we use a numerical approximation of the energy spectrum (\cref{Eq: Energy Spec def2}) with $\mathcal{T}=1/\xi$ to simulate the limited observable accuracy $\xi$ and differentiate from the exact energy spectrum in \cref{Fig: Pert Resonance1}.
	\begin{equation}
		P^{\mathcal{T}}(E)=\abs{\frac{1}{2\pi}\int\limits_{-\mathcal{T}}^{+\mathcal{T}}e^{iEt}\ket{\Psi(t)}\dd{t}}.
	\end{equation}
	
	Including the finite accuracy effect on the energy spectrum, we find it to closely resemble the unperturbed spectrum \cref{Fig: Resonance1}. The stationary point of the effective average energy (\cref{Eq: Two-level effective average}) as well as that of the spectrum average taken from \cref{Fig: Pert Resonance2}, closely agree with the unperturbed solution. We depict the acceptable accuracy range calculated from \cref{Eq: Acceptable accuracy range}, as the vertical gray areas, with the inner most area corresponding to an error of $v$ instead of $\xi$. The error in the eigenstate wavefunctions is normalized by the average energy difference as follows:
	\begin{equation}
		\ket{\Psi_\pm(0)}=\ket{\Psi^0_\pm(0)}+\order{\frac{\xi}{\bar{\epsilon}^0_+-\bar{\epsilon}^0_-}}=\ket{\Psi^0_\pm(0)}+\order{\frac{\xi}{\omega_0}}\label{Eq: Acceptable accuracy range}.
	\end{equation}
	
	Finally we show the possibility of the Ritz variation principle by adapting Davidson algorithm to the Floquet system, similar to Sahoo et. al. \cite{Sahoo_2019}. This example system is simple enough that we can use this formulation directly, while more complicated systems would require a more refined method to take into account the average energy variation as well. Starting from the initial guess of the undriven ground state $u^{i_0}$, we calculate the ground states $u^{i_n}$ as defined in this paper and the residue vector $r^{i_n}$ at each step $i_n$, which represent the current approximation of the ground state and the convergence condition, respectively. We have executed the procedure equivalent to \cite{Sahoo_2019} and we will not go into detail about this procedure. We only note that we've changed the algorithm of selecting the ground state and we focus on how the algorithm converges and the corresponding solution.
	\begin{align}
		\ket{u^{i_0}(t)}=&\mqty(0\\1)&\ket{r^{i_0}(t)}=&\mqty(v+\frac{V}{2}e^{-i\omega t}\\0)\quad\gg\xi,\\
		\ket{u^{i_1}(t)}=&\ket{u^{0\infty}_+(t)}+\order{v}&\ket{r^{i_1}(t)}=&\order{v}\quad<\xi.
	\end{align}
	The results converge to an error of $\xi$ within two steps, without having to create the infinite matrix $H^F$. The convergent solution is also well within the accepted accuracy to the observed ground state. If we disregard the acceptable accuracy $\xi$ and instead choose an accuracy $\ll v$, we would eventually converge to the exact infinite time ground state.
	
\section{Conclusion}
	In this paper we have proposed a robust method of defining the Floquet eigenstate in an ordered fashion using the average energy. The main differences from the previous methods are: we are able to uniquely order the eigenstates independent of the quasi-energy shift; we can uniquely define the eigenstates near resonance where it would otherwise be ambiguous; the method is robust against infinitesimal perturbations; we can systematically cut off the Hilbert space and retain the accuracy within a reasonable timescale. Based on this definition we derived variational methods of approximating the eigenstates, which in principle are more computationally efficient. For this method additional consideration has to be made to the physical timescale that we wish to investigate.
	
	The physical significance of the average energy was not presented in this current work. Some intuition can be found by evaluating the steady state of the open-quantum system. We conjecture that for reasonable systems, this choice of basis set gives a good approximation of the exact steady state, i.e. the density matrix is close to diagonal and occupied by a few low lying states in this eigenstate representation, even as the Hilbert space is truncated.
	
	Having a variational principle for deriving the ground state, various method such as Floquet Hartree-Fock could be adapted to approximate the ground state. Since the ground state is generally not sufficient to describe a physical steady-state, additional excited states are required, which can be derived in a similar variational manner, with similar Excited Floquet Hartree-Fock. With sufficient calculations (estimated from the resulting energy spectrum), the computation of the physical Floquet steady-state is relatively trivial, at which point we have a good approximation of a real periodically driven system at a long enough time-scale to have equilibrated with the environment.
	
\begin{acknowledgments}
	This work was supported in part by the Innovative Center for Coherent Photon Technology (ICCPT) in Japan and the Center of Innovation Program from the Japan Science and Technology Agency, JST. C.M.L. was supported by the Japan Society for the Promotion of Science through the Program for Leading Graduate Schools (MERIT) and Professional  development  Consortium  Computational  Material Scientists (PCoMS). C.M.L. would also like to thank Professor Peter Maksym for the insightful discussions and general support.
\end{acknowledgments}
	\appendix
	\section{Observed average energy boundaries\label{Sec: Average energy boundary}}
	We can find the exact boundaries $T^{\bar{E}}_{min}$ and $T^{\bar{E}}_{max}$ where the observed average energy functional of a model/perturbed Hamiltonian $H(t)$ approximates that of the unperturbed Hamiltonian $H^0(t)$, by decomposing \cref{Eq: Average energy condition} onto the unperturbed eigenbasis $\mathcal{E}_{H^0}$, and imposing the condition in \cref{Eq: Average energy condition2}.
	\begin{widetext}
	\begin{align}
		\bar{E}^\mathcal{T}=&\sum_{n}\abs{C^0_n}^2\frac{1}{\mathcal{T}}\int_{0}^{\mathcal{T}}\ev*{\hat{H}^0(t)}{u^0_n(t)}\dd{t}\beginsubeqn\label{Eq: Approx average energy expansion0}\\
		&+\sum_{m\neq n}C^{0*}_mC^0_n\frac{1}{\mathcal{T}}\int_{0}^{\mathcal{T}}e^{-i(\epsilon^0_n-\epsilon^0_m)t}\mel*{u^0_m(t)}{i\partial_t}{u^0_n(t)}\dd{t}\subeqn\label{Eq: Approx average energy expansion1}\\
		&+\sum_{m,n}C^{0*}_mC^0_n\frac{1}{\mathcal{T}}\int_{0}^{\mathcal{T}}e^{-i(\epsilon^0_n-\epsilon^0_m)t}\mel*{u_m(t)}{[\hat{v}(t)-i\int_{0}^{t}\comm{\hat{H}^0(t)}{\hat{v}_I(\tau)}\dd{\tau}}{u_n(t)}\dd{t}+\order{v^2}\subeqn\label{Eq: Approx average energy expansion2}.
	\end{align}
	\end{widetext}
	
	The first term (\cref{Eq: Approx average energy expansion0}) quickly converges to the unperturbed average energy $\bar{E}^0[\Psi]$ (regardless of its definition) within a few cycles of the driving $\mathcal{T}>\order{T}$. This timescale is much lower than the timescales we will be discussing and will be ignored. Thus the goal is to find the timescale $\mathcal{T}$ where the remaining terms (\cref{Eq: Approx average energy expansion1,Eq: Approx average energy expansion2}) vanish for arbitrary initial wavefunction $\Psi(0)$, or in this experssion, arbitrary coefficients $\qty{C^0_n}$.
	
	From the second term (\cref{Eq: Approx average energy expansion1}), we obtain the lower boundary $T^{\bar{E}}_{min}$ which is independent of the perturbation $v$ of the model $H(t)$. Depending on which definition of the average energy and eigenbasis we wish to approximate in the unperturbed system (\cref{Sec: Real near resonance}), the lower boundary $T^{\bar{E}}_{min}$ can change. For simplicity we assume the unperturbed system has no finite near resonances, so that we approximate the infinite time solutions of the unperturbed system $\bar{E}^{\mathcal{T}}[\Psi]\approx\bar{E}^{0\infty}[\Psi]$. The lower boundary is thus obtained from:
	\begin{gather}
		\frac{e^{-i\omega_{mnl}^0\mathcal{T}}-1}{\omega_{mnl}^0\mathcal{T}}\sum_{k}k\omega\ip*{u_m^{0(k)}}{u_n^{0(k+l)}}<\xi\qquad\forall\; m,n,l,\\
		T^{\bar{E}}_{min}=\max_{m\neq n,l}\abs{\frac{2\sum_{k}k\omega\ip*{u_m^{0(k)}}{u_n^{0(k+l)}}}{\xi\omega_{mnl}^0}}\label{Eq: Def min boundary},\\
		\bar{E}^{0\mathcal{T}}=\sum_{n}\abs{C^0_n}^2\bar{\epsilon}^{0\infty}_n+\order{\xi}\qquad\forall\; \mathcal{T}>T^{\bar{E}}_{min}.
	\end{gather}
	Even when the unperturbed Hamiltonian has exact resonances ($\exists\,\omega^0_{mnl}=0$), the lower boundary remains finite due to the diagonalized definition of the infinite time average energy (\cref{Eq: Diagonal inf average energy}) and the infinite time eigenstate definition. Similarly in the case of infinitesimally small, but finite near resonance ($\exists\,\abs{\omega^0_{mnl}}\ll\xi$), the lower boundary can be within the acceptable timescale $T^{\bar{E}}_{min}<\mathcal{T}\sim1/\xi$, depending on whether we treat these states as resonant states or not (\cref{Sec: Real near resonance}). We can further lower this boundary if we limit the average energy functional approximation to a smaller Hilbert space, e.g. only approximating the functional near the ground state. A rough physical understanding of the boundary $T^{\bar{E}}_{min}$ is the timescale from where the weakest significant interaction or avoided crossing can be resolved up to the accuracy $\xi$.
	
	The higher boundary $T^{\bar{E}}_{max}$ is derived from the remaining terms in \cref{Eq: Approx average energy expansion2}, and are specific to each perturbation $v$. We can safely ignore the perturbation effects on non-resonant states and only consider the effects of resonant and near-resonant states ($\text{mod}(\epsilon_a-\epsilon_b,\omega)<\xi$). Calculating the condition for \cref{Eq: Average energy condition2} to be satisfied we get a form of the higher boundary as follows:
	\begin{gather}
		\frac{\mathcal{T}^2(\bar{\epsilon}^{0\infty}_a-\bar{\epsilon}^{0\infty}_b)v_{ab}^{0(0)}}{2\mathcal{T}}<\xi\qquad\forall a,b,\\
		T^{\bar{E}}_{max}=\min_{a\neq b}\abs{\frac{2\xi}{(\bar{\epsilon}^{0\infty}_a-\bar{\epsilon}^{0\infty}_b)v_{ab}^{0(0)}}}\label{Eq: Def max boundary},\\
		\bar{E}^{\mathcal{T}}=\sum_{n}\abs{C^0_n}^2\bar{\epsilon}^{0\infty}_n+\order{\xi}\qquad\forall\; T^{\bar{E}}_{min}<\mathcal{T}<T^{\bar{E}}_{max}.
	\end{gather}
	
	This higher boundary $T^{\bar{E}}_{max}$ closely approximates $T^U_{max}$ which defines up to which timescale the propagator and any observable, are closely approximated in the model/perturbed system and the real/unperturbed one. So a rough physical intuition of this higher boundary is the timescale up to which the effects of the infinitesimal resonance lifting can be ignored.
	
	Depending on what we define to be the perturbation $v$, and what the unperturbed eigenstates are, the lower and higher boundaries could cross ($T^{\bar{E}}_{max}<T^{\bar{E}}_{min}$), in which case we would not be able to find an average energy definition that satisfies $\abs{\bar{E}^{\mathcal{T}}-\bar{E}^{0\mathcal{T}}}<\xi$, and the Floquet eigenstates of the two systems could differ significantly for any observed average energy definition. Different choices of perturbation $v$ or unperturbed eigenstate defintion $\mathcal{E}^{\mathcal{T}}_{H^0}$ could recover this condition (e.g. \cref{Sec: Real near resonance}). Otherwise it could simply be that the perturbation $v$ is not weak enough to be ignored and the model does not closely approximate the real system.
	
	\section{Real average energy near resonance\label{Sec: Real near resonance}}
	In \cref{Sec: Approximate average energy} we have assumed that the real system does not have near resonance conditions so that the infinite time eigenstates are the optimal basis set to describe the system at reasonable timescales $\mathcal{T}<T^U_{max}$. But the real system can have infinitesimal near resonance like the model system, in which case the procedure presented in \cref{Sec: Approximate average energy} would not approximate the infinite time eigenstates of the real system.
	\begin{equation}
		\mathcal{E}^{\mathcal{T}}_{H^0}\napprox\mathcal{E}^{\infty}_{H^0}\quad if\quad\exists\; 0<\text{mod}(\epsilon^0_n-\epsilon^0_m,\omega)<\xi.
	\end{equation}
	
	We now consider the weak interaction $v'(t)$ which would bring these near resonance pairs to exact resonance, so that we have a Hamiltonian $H'(t)$ with only exactly resonant and far from resonant Floquet eigenstates which can be derived from the procedures in \cref{Sec: Exact average energy}.
	\begin{gather}
		H'(t)=H^0(t)+v'(t),\\
		\mod(\epsilon'_n-\epsilon'_m,\omega)\in\qty{0,\gg\xi}\qquad\forall\;m,n.
	\end{gather}
	
	Repeating the procedures in \cref{Sec: Approximate average energy,Sec: Average energy boundary}, with an exchange of Hamiltonians, we can find the timescale boundaries $T^{\bar{E}}_{min}$ and $ T^{\prime\bar{E}}_{max}$ where the observed average energy functional and its eigenstates (derived from \cref{Eq: Variational method practical0,Eq: Variational method practical1,Eq: Variational method practical2}) approximate the infinite time solutions of this model Hamiltonian $H'(t)$.
	\begin{gather}
		H(t)\to H^0(t)\qquad H^0(t)\to H'(t),\\
		\abs{\bar{E}^{0\mathcal{T}}[\Psi]-\bar{E}^{\prime\infty}[\Psi]}<\xi\qquad\forall\;\Psi(0);\;\mathcal{T}\in\qty\big[T^{\bar{E}}_{min},T^{\prime\bar{E}}_{max}],\\
		\bar{\epsilon}^{0\mathcal{T}}_n\approx\bar{\epsilon}^{\prime\infty}_n+\order{\xi}\qquad \Psi^0_n\approx\Psi'_n+\order{\xi}.
	\end{gather}
	In this case the lower boundary $T^{\bar{E}}_{min}$ can be within acceptable timescales $T^{\bar{E}}_{min}<\mathcal{T}\sim1/\xi<T^U_{max}$. This is in contrast with the lower boundary $T^{\bar{E}\infty}_{min}$ required to resolve the infinitesimal, but finite near resonance/coupling $v'$ in the real system $H^0(t)$. For small enough interaction $\abs{v'}\ll\xi$, the upper boundary timescale is beyond experimental observations $1/\xi<T^U_{max}\ll T^{\prime\bar{E}}_{max}$, and we can define the observed eigenstates of the real system $\mathcal{E}^\mathcal{T}_{H^0}=\qty{\epsilon^0_n,\bar{\epsilon}^{0\mathcal{T}}_n,u^0_n(t)}$. By following \cref{Eq: Variational method practical0,Eq: Variational method practical1,Eq: Variational method practical2}, we do not even have to find the model Hamiltonian $H'(t)$ which it approximates, and calculate the observed eigenstates directly.
	
	From the original discussion in \cref{Sec: Approximate average energy}, we have another higher bound $T^{\bar{E}}_{max}$, beyond which the observed solutions $\mathcal{E}^{\mathcal{T}}_{H}$ of the model Hamiltonian $H(t)$ differ from the real ones $\mathcal{E}^{\mathcal{T}}_{H^0}$ of $H^0(t)$. In principle the infinitesimal perturbation/numerical errors $v$ in the model Hamiltonian $H(t)$ would be larger than the interaction $v'$. So in principle we do not need to consider the higher boundaries of the real system $T^{\bar{E}}_{max}<T^{\prime\bar{E}}_{max}$, and the observed eigenstate solutions of the model $H(t)$ are guaranteed to correspond to the ones of the real system $H^0(t)$ for $\mathcal{T}\in[T^{\bar{E}}_{min};T^{\bar{E}}_{max}]$.
\bibliography{Floquet-Variation.bib}
\end{document}